\documentclass[12pt,a4paper]{article}
\usepackage{rotating}
\usepackage{amstext}
\usepackage{amssymb}
\def\be{\begin{equation}}
\def\ee{\end{equation}}
\def\bq{\begin{eqnarray}}
\def\eq{\end{eqnarray}}
\def\beq{\begin{eqnarray}}
\def\eeq{\end{eqnarray}}
\def\ov{\overline}
\def\un{\underline}

\def\epl{e^{(\lambda)}}
\newcommand{\sect}[1]{\setcounter{equation}{0}\section{#1}}

\newcommand{\hs}{h_\parallel^{(s)}}

\newcommand{\htt}{h_\parallel^{(t)}}

\newcommand{\f}{f_\rho}
\newcommand{\m}{m_\rho}

\def\mx{\frac{x^2m^2_\rho}{4}}

\def\Drl{\stackrel{\leftrightarrow}{D}}
\def\Dl{\stackrel{\leftarrow}{D}}
\def\Dr{\stackrel{\rightarrow}{D}}
\def\ud{\uparrow\downarrow}
\def\du{\downarrow\uparrow}

\def\ubar{\overline{u}}

\def\Gtilde{\tilde{G}}

\begin{document}

\font\fifteen=cmbx10 at 15pt
\font\twelve=cmbx10 at 12pt

\begin{titlepage}

\begin{center}

\renewcommand{\thefootnote}{\fnsymbol{footnote}}

{\twelve Centre de Physique Th\'eorique\footnote{Unit\'e Propre de
Recherche 7061 }, CNRS Luminy, Case 907}

{\twelve F-13288 Marseille -- Cedex 9}

\vspace{1 cm}

{\fifteen Light-cone wave functions of mesons}

\vspace{0.3 cm}

\setcounter{footnote}{0}
\renewcommand{\thefootnote}{\arabic{footnote}}

{\bf Gautier STOLL}

\vspace{2,3 cm}

{\bf Abstract}

\end{center}

A review of light-cone, covariant and gauge-invariant wave-functions of 
mesons is presented. They 
are basic non-perturbative objects needed for hard exclusive processes and for the method of light-cone QCD sum rules. The emphasis is on the vector mesons and a new (model-independent) way of computing the mass corrections of vector-meson wave functions is given.

\vspace{2 cm}

\noindent Key-Words: Non-perturbative QCD, Conformal expansion.

\bigskip

\noindent December 1998

\noindent CPT-98/P.3735

\bigskip

\noindent anonymous ftp : ftp.cpt.univ-mrs.fr

\noindent web : www.cpt.univ-mrs.fr

\renewcommand{\thefootnote}{\fnsymbol{footnote}}

\end{titlepage}

\setcounter{footnote}{0}
\renewcommand{\thefootnote}{\arabic{footnote}}

\sect{Introduction}

One of the main goal in QCD is to understand the non-perturbative aspects of this theory. In this report, I describe a way of parameterizing these non-perturbative effects in a covariant and gauge-invariant way : the "meson wave functions" or "light-cone distributions amplitudes of meson", with the emphasis on vector meson (the pseudoscalar case, like the $\pi$-meson, is simpler and the results can be found in \cite{BF}). By definition, a "wave function" is the matrix element of (gluon and) quark operators on the light-cone between the vacuum and a light meson. It is conceptually  different to the OPE applied to QCD (like in the method of QCD sum rules \cite{SVZ}) where the non-perturbative inputs are vacuum expectation values of {\sl local} operators.

The standard approach to wave function (due to Brodsky and Lepage\cite{BrLe, BrLe2}) considers the parton decomposition in the infinite momentum frame. A mathematically equivalent formalism is the light-cone quantization\cite{BrPa}. Here, the concept of wave function is defined in a different way. Although it is less intuitive, it keeps the Lorentz and gauge invariance, in order to use the equation of motion of the QCD.With this approach, one gets directly the non-perturbative objects needed for the method of light-cone QCD sum rules\cite{B}, which allows to compute exclusive decays of heavy hadron beyond perturbative QCD. 

To compute a whole non-perturbative function seems an impossible task. Therefore the "conformal expansion" is used (it is like a partial-wave expansion, but referred to the conformal invariance of QCD), and the unknown input come from local matrix elements, which can be computed with the help of QCD sum rules\cite{SVZ}. Within this expansion the renormalization-scale dependence is also under control.

Like in DIS, the different wave functions are classified by their increasing twist. In order to be more precise when using these wave functions, the non-leading twist ones are needed. A part of these non-leading twist contributions comes from the mass correction, which can be important if $K^*$ is considered for example.

In this work, after two sections of preliminaries and definitions, I present the conformal expansion of these wave functions : I introduce the conformal group, the notion of conformal operators on the light-cone, the link between conformal invariance and renormalization, the matrix elements of conformal operators. Then I give an example of computation of non-leading twist three-point wave function. Finally, I give methods of computing mass corrections, with some new formulas which gives the mass corrections directly from the leading twist wave function (without making a conformal expansion). I add some appendices which contain lists of wave functions and technical results.

%
%

\sect{General framework}
The two basic objects are (for the $\rho$-meson)~:

\noindent 1) The two-point wave functions, extracted from this kind of matrix element~:
\be
\left\langle 0 |\ov{u}(x)[x,-x] \Gamma d(-x)|\rho \right\rangle \label{2matel}
\ee

\noindent 2) The three-point wave functions, extracted from this kind of matrix element :
\be
\left\langle 0 | \ov{u}(x)[x,vx] g G_{\mu\nu}(vx)[vx,-x] \Gamma d(-x)|\rho\right\rangle
\label{3matel}
\ee
Where $x$ is almost on the light cone, $v \in [0,1]$, $\Gamma$ any kind of product of $\gamma_{\mu}$ matrices and $[x,y]$ a path-ordered gauge factor along the straight line connecting $x$ and $y$ :
\be
[x,y]={\rm P}\exp \left[ ig\int^1_0 dt (x-y)_\mu A^\mu (tx+(1-t)y)\right]
\ee

This gauge factor is a way to introduce interaction (see \cite{Bali}), it ensure the gauge invariance of these non-local matrix elements (I will omit to write it sometimes).

These matrix elements depend on three vectors : 

\noindent - $P_{\mu}$ : momentum of the $\rho$-meson

\noindent - $\epl_{\mu}$ : polarization vector of the $\rho$-meson

\noindent - $x_{\mu}$

With the relations :
\bq 
P^2&=&m^2_{\rho} \\ \epl\cdot\epl&=&-1 \\ P\cdot\epl&=&0
\eq

The parameterization of these different matrix elements is based on the Operator Product Expansion on the light-cone\cite{BP}. So one needs light-like vectors $p$ and $z$, built from $P$, $\epl$ and $x$ :

\bq
p^2&=&0 \\ z^2&=&0
\eq
such that $p\rightarrow P$ in the limit $m^2_{\rho} \rightarrow 0$ and $z\rightarrow x$ for $x^2\rightarrow 0$ (see \cite{BBKT}) :
\bq
z_\mu &=& x_\mu - P_\mu \frac{1}{m^2_\rho}\left[xP-\sqrt{\left(xP\right)^2 - x^2m^2_\rho}\right] \label {defz} \\
p_\mu &=& P_\mu - \frac{1}{2} z_\mu\frac{m^2_\rho}{pz}. \label{defp}
\eq

Useful scalar products are
\bq
zP=zp&=&\sqrt{\left(xP\right)^2 - x^2m^2_\rho} \\
p\cdot\epl&=&-\frac{m^2_\rho}{2pz}z\cdot\epl.
\eq

The polarization vector $\epl$ can be decomposed in projections onto the two light-like vectors and the orthogonal plane :
\be
\epl_\mu=\frac{(\epl\cdot z)}{pz}\left(p_\mu -\frac{m^2_\rho}{2pz}z_\mu\right)+\epl_{\perp\mu}. \label{defeperp}
\ee

Note that
\be
(\epl\cdot z)=(\epl\cdot x).
\ee

For a practical use, one defines the projector onto the direction orthogonal to $p,z$ :
\be
g^{\perp}_{\mu\nu}=g_{\mu\nu}-\frac{1}{pz}(p_\mu z_\nu + p_\nu z_\mu), \label{gperp}
\ee
and these notations :
\bq 
a.&\equiv & a_\mu z^\mu \nonumber \\ a_* & \equiv & a_\mu p^\mu / (pz) \label{dotstar}
\eq

%
%

\sect{Definitions of twist \label{deftw}}

The different contributions to the matrix elements \ref{2matel}, \ref{3matel} are classified by their "twist", but there are two different
definitions of this concept : 

For local operators, the "twist" means dimension minus Lorentz-spin. To extract the contribution of a definite twist for non-local matrix elements like \ref{2matel}, \ref{3matel}, an expansion in local operator (OPE) has to be done and all local operators which have a definite twist in the sense given above must be re-summed. This will give
the non-local term which has a definite power in $x^2$. To understand how this
definition of "twist" means "power in $x^2$", see \cite{BP}. From now on, this definition will be called "theoretical twist".

The second definition is built in analogy to partons distributions. In the latter case, the different structures of non-local matrix elements are separated with their different powers of $Q$ (the hard momentum transfer) in deep inelastic scattering : a term of
twist $t$ contributes to the inclusive cross section with coefficients which contain $t-2$ or more power of $1/Q$. A good description of this classification can be found in \cite{JJ}. For the matrix elements \ref{2matel}, \ref{3matel}, the same type of definition can be used, like in \cite{BBKT} : structures with $p\cdot z$, with $p_\mu$ or with $\epl\cdot z$ gives one power of $Q$, but $\epl_\perp$ behaves as $Q^0$. From now on, this definition will be called "physical twist".

As an example, one gives the parameterization of a two-point matrix element :

\bq
\lefteqn{\langle 0|\ov{u} (z) \gamma_\mu [z,-z] d ( -z)|\rho^- (P, \lambda) \rangle 
 = }\nonumber\\
&=& f_\rho m_\rho\left[ p_\mu \frac{\epl\cdot z}{p\cdot z}\right. \int^1_0 du e^{i\xi p 
\cdot z}\phi_\| (u,\mu^2) + \epl_\perp\int^1_0 du e^{i\xi p \cdot z}
g_\perp^{(v)}(u,\mu^2)  \nonumber\\
& &{}\left.-\frac{1}{2}z_\mu\frac{\epl\cdot z}{(p\cdot z)^2} m^2_\rho \int^1_0 du e^{i\xi
p\cdot z} g_3(u,\mu^2)\right]
\eq
where
\be
\xi=2u-1,
\ee
$\mu$ is the renormalization scale and
$f_\rho$ is defined by the following local matrix element :
\be
\langle 0|\ov{u} (0)\gamma_\mu d(0)|\rho^-(P,\lambda)\rangle = f_\rho m_\rho
\epl_\mu.
\ee

All the three functions $\phi_\|,g_\perp^{(v)},g_3$ are normalized like this :
\be
\int^1_0 du \phi_\| (u)=1
\ee

Following the classification of \cite{BBKT}, $\phi_\|$ is a ("physical"-)twist 2 contribution, $g_\perp^{(v)}$ is ("physical"-)twist 3 and $g_3$ is ("physical"-)twist 4. But when one looks at the definition
given above, one would argue that these contributions should be twist 1, 2, 3. In fact, the authors of \cite{BBKT} used another definition of twist, which just
shifts it to one unity, based on the light-cone quantization formalism. The parameterization of all matrix elements \ref{2matel} and \ref{2matel} up to twist four can be found in the first appendix.

A physical meaning for the different functions  $\phi_\|, g_\perp^{(v)}, g_3$ can be given : they describe the probability amplitude to find the $\rho$ in a state with a quark and an antiquark which carry momentum fractions $u$ for the quark and $1-u$ for the antiquark respectively and a small transverse separation of order $1/\mu$.

Roughly speaking, one uses the "physical twist" to define the different wave functions. But in order to compute them, an expansion in "theoretical twist" is made.

%
%

\sect{Conformal expansion}

When the different wave functions which can be extracted from
non-local matrix elements like \ref{2matel} and \ref{3matel} has been classified, one needs to compute them. The first idea is to expand them in the different moments, like in \cite{CZ}. But it is like doing an expansion in local operators (OPE). This must be avoided if these wave functions are needed for a light-cone QCD sum rule (see \cite{B}). 

The method used to compute wave functions is the "conformal expansion" : the idea is to expand the wave function in a series of polynomials, whose coefficients are renormalized (almost) multiplicatively. Here is an example for $\phi_\|$ (with $\xi=2u-1$) :
\be
\phi_\| (u)=6u\ubar \sum^{\infty}_{n=0}a^\|_nC_n^{3/2} (\xi),
\xi=2u-1 \label{confex}
\ee
where $C_n^{3/2}$ is the Gegenbauer polynomial (see appendix \ref{app:a}) and $a_n^\|$ are coefficients which are renormalized multiplicatively (to one loop). These $a_n^\|$ can be computed using the usual QCD sum rules and exact relations (see section \ref{PsiPhi}). 

In this section, the method of building such an expansion will be described. For it, one needs to know what is the structure of the conformal group, in what sense is QCD conformal invariant, how to build local and polylocal conformal operators and
finally what happens for the matrix element of polylocal conformal operators.

%

\subsection{Structure of the conformal group}

The conformal group is a Lie group which contains the Poincar\'e group and the dilatation. To describe its structure, it is simpler to work directly with the algebra. The generators of the conformal group are the generators of the Poincar\'e algebra $P_\mu$ and $M_{\mu\nu}$, the dilatation $D$ and of the special conformal transformations $K_\mu$.

They satisfy the following commutation relations\cite{BF,Oh,MS} :
\bq
[D,K_\mu] & = & iK_\mu \\
{}[ K_\mu ,P_\nu ] & = & -2i (g_{\mu\nu}D+M_{\mu\nu}) \\
{}[D,M_{\mu\nu}]&=&0 \\
{}[D,P_\mu]&=&-iP_\mu \\
{}[K_\rho,M_{\mu\nu}]&=&i(g_{\rho\mu}K_\nu-g_{\rho\nu}K_\mu) \\
{}[K_\mu,K_\nu]&=&0
\eq
plus those of the Poincar\'e algebra. The action of these generators on a field $\Phi(x)$ with an arbitrary spin and canonical dimension $l$ is the following :
\bq
[P_\mu,\Phi(x)]&=&-i\partial_\mu\Phi(x) \label{conf1} \\
{}[M_{\mu\nu},\Phi(x)]&=&[i(x_\nu\partial_\mu-x_\mu\partial_\nu)-\Sigma_{\mu\nu}]
\Phi(x) \\
{}[D,\Phi(x)&=&-i(x^\xi\partial_\xi+l)\Phi(x) \\
{}[K_\mu,\Phi(x)]&=&-i(2x_\mu x^\xi\partial_\xi-x^2\partial_\mu+2x_\mu l-2ix_\nu\Sigma_{\mu\nu})\Phi(x) \label{conf4}
\eq
where $\Sigma_{\mu\nu}$ is the generator of spin. It acts on a fermion field $\psi$ and a gauge field $G_{\alpha\beta}$ in the following way :
\bq
\Sigma_{\mu\nu}\psi&=&\frac{1}{2}\sigma_{\mu\nu}\psi \label{fspin} \\
\Sigma_{\mu\nu}G_{\alpha\beta}&=&i(g_{\mu\alpha}G_{\nu\beta}-g_{\nu\alpha}
G_{\mu\beta})-(\alpha \leftrightarrow \beta) \label{gspin}
\eq

%

\subsection{Conformal invariance and QCD \label{sec:confQCD}}

Anyone can convince himself that the unrenormalized QCD-action (with massless
quark) is conformal
invariant. But this symmetry is broken by two effects :

\noindent 1) Renormalization

\noindent 2) Quark-mass terms

But a careful analysis of these effects shows that the conformal expansion of wave functions like in \ref{confex} is still relevant.

\subsubsection{Renormalization effect on the conformal invariance}

Suppose one has a conformal operator in non-interacting and massless QCD~: it is an operator which does not mix with other ones under conformal transformation (such local and polylocal operators are constructed in the next subsection). Following \cite{Ma}, one can prove that such operators do not mix under renormalization at the leading log approximation (if simple derivatives are transformed into covariant ones). It means that conformal symmetry (and conformal expansion for wave function) is still a "good" symmetry of massless QCD.

Consider an operator which renormalizes multiplicatively. Its matrix \mbox{elements} $\Gamma$ satisfies the Callan-Symanzik equation :
\be
\left[\mu\frac{\partial}{\partial\mu}+\beta (g)\frac{\partial}{\partial g}\right]
\Gamma(p,\mu;g)=\gamma_\Gamma (g)\Gamma(p,\mu;g). \label{CaSy}
\ee
where $p\equiv (p^1 _\mu, \ldots ,p^N_\mu)$ is a set of momenta and $\gamma_\Gamma$ is the anomalous dimension of the operator considered. If this matrix element comes from a massless theory, it is an homogeneous function of $p$ and $\mu$. The Euler's theorem yields :
\be 
\left[\mu\frac{\partial}{\partial\mu}+\rho\frac{\partial}{\partial\rho}\right]
\Gamma(\rho p,\mu;g)=d_\Gamma\Gamma(\rho p,\mu;g),
\ee
where $d_\Gamma$ is the canonical dimension of the operator considered. So equation \ref{CaSy} can be transformed in
\be
\left[\rho\frac{\partial}{\partial\rho}-\beta (g)\frac{\partial}{\partial g}
\right]
\Gamma(\rho p,\mu;g)=\left[d_\Gamma-\gamma_\Gamma(g)\right]\Gamma(\rho p,\mu;g).
\ee

If one considers the theory at one loop level, $\beta(g)$ can be neglected because it starts with $g^3$; one gets :
\be
\rho\frac{\partial}{\partial\rho}
\Gamma(\rho p,\mu;g)=\left[d_\Gamma-\gamma_\Gamma(g)\right]\Gamma(\rho p,\mu;g).
\ee

This last equation shows that $\Gamma$ satisfies the dilatational Ward identity, so if the interaction is switched off, one gets a matrix element which transforms in itself under a dilatation~: it is the matrix element of a conformal
operator.

So if an operator is renormalized multiplicatively at one loop order, it is conformal if the interaction is switched off. Is the converse true ?

Suppose one has a set of operators. A basis which diagonalizes the anomalous dimension matrix at one loop order is built. From the preceding arguments, one knows that this basis is made of conformal operators. If one directly builds a basis of conformal operators (like in the next section), almost the same basis of multiplicatively renormalized operator is obtained, except if the eigenvalues of the anomalous matrix (or in the representation of the conformal group) are degenerate.

So conformal operators renormalize almost multiplicatively at one loop order, they can only mix with the ones which have the same conformal representation. The
coefficient of the dilatational Ward identity transforms like this
\be
d_\Gamma\rightarrow d_\Gamma-g^2\gamma_2
\ee 
when the interaction is switched on at one loop order. $g^2\gamma_2$ is the value of the anomalous dimension $\gamma_\Gamma(g)$ at the order $g^2$ :
\be
\gamma_\Gamma(g)=g^2\gamma_2+g^4\gamma_4+\ldots
\ee

This property of conformal operators works also in the leading logarithmic approximation, otherwise it would contradict this property at one loop order (see \cite{Ma}).

In \cite{Oh}, the anomalous dimension of some conformal operators is computed and for these operators, it is checked that they renormalized multiplicatively to one loop.

Some complications arise when one deals with gauge theories. In order to have gauge-invariant operators, usual derivatives have to be replaced by covariant ones. But in that case, it is not sure that the diagonalization of the anomalous dimensions matrix is really possible. But one hopes that for color-singlet operators, it works (see \cite{Oh}). The other problem is that an operator which transforms in itself under a dilatation may not be a conformal one (the action of
special conformal transformations $K_\mu$ may introduce some gauge-variant terms). But it seems very hard to construct such "almost"-conformal operators, so it is assumed that they do not exist (see \cite{Sa}).

The coefficients of the conformal expansion of a wave function (like the $a_n^\|$ in equation \ref{confex}) are matrix elements of local conformal operators (see sections \ref{sec:locconf} and \ref{PsiPhi}). So they are renormalized multiplicatively, that is why conformal expansion like in \ref{confex} make sense.

\subsubsection{Quark-mass effect on the conformal invariance}

Masses of light-quarks break explicitly the conformal invariance of the QCD-Lagrangian. So they can break the multiplicative renormalizability of conformal
operators. But if the renormalization scale is high enough compare to the 
light quark masses, this effect is small and the conformal expansion of wave functions is still a "good" expansion.

Quark masses can also appear explicitly
in the twist-expansion of matrix elements of non-local operators.  It that case, it introduce some new wave functions, but the latter can also be expanded in a conformal way.

So the effect of quark masses can be controlled and the conformal expansion is
still a good way to compute the wave functions.

%

\subsection{Construction of local and polylocal conformal operators}

\subsubsection{Local operators on the light-cone \label{sec:locconf}}

For the use of QCD light-cone sum rules, one deals with fields $\Phi(z)$ varying on the light cone ($z^2=0$) or almost on the light-cone. But the representation of the conformal group is much simpler on the light cone, that is why a light-like
basis $z_\mu$, $p_\mu$ was introduced at the beginning of this chapter (equations \ref{defz} and \ref{defp}). 

Following \cite{BF}, one can see that only the components $P.$, $D$, $M_*.$ and
$K_*$ of the algebra of the conformal group only act non-trivially on the field $\Phi(z)$ (the meaning of $.$ and ${}_*$ is given at equation \ref{dotstar}). These generators form the algebra of a subgroup called collinear conformal subgroup $SO(2,1)\cong SU(1,1)\cong SL_2(R)$ which generates projective transformations on the line (see e. g. \cite{Oh}).

This algebra can be brought into  a more standard form, with the following linear combinations :

\bq
J_+=J_1+iJ_2&=&\frac{i}{\sqrt{2}}P. \\
J_-=J_1-iJ_2&=&\frac{i}{\sqrt{2}}K_* \\
J_3&=&\frac{i}{\sqrt{2}}(D+M_*.) \\
E&=&\frac{i}{\sqrt{2}}(D-M_*.) \\
J^2&=&J_3^2-J_1^2-J_2^2=J_3^2-J_3-J_+J_-.
\eq

Then the commutation relations are
\bq
[J_+,J_-]&=&-2J_3 \\
{}[J_3,J_\pm ]&=&\pm J_\pm \\
{}[E,J_i]&=&0 \\
{}[J^2,J_i]&=&0
\eq

In order to build a representation of this subgroup, all the generators  $P.$, $D$, $M_*.$ and $K_*$ must reduce to differential operators. Equations \ref{conf1} to \ref{conf4} imply that fields must be eigenvectors of the operator $\Sigma_*.$, that is fields having fixed projections ($s$) of the Lorentz spin on to the line $z_\mu$ :
\be
\Sigma_*.\Phi(z)=is\Phi(z)
\ee

A spinor field $\psi$ has two components with the projection $s=\pm \frac{1}{2}$ :
$\gamma .\psi$ and $\gamma_*\psi$. One can see it using equation \ref{fspin} :

\bq
\Sigma_*.\gamma.\psi=\frac{1}{2}\sigma_*.\gamma.\psi=-\frac{i}{2}\gamma.\psi 
\label{confop1} \\
\Sigma_*.\gamma_*\psi=\frac{1}{2}\sigma_*.\gamma_*\psi=+\frac{i}{2}\gamma_*\psi
\eq

For the gauge field $G_{\mu\nu}$ there are three possible value of the projection : $s=-1,0,1$. The different components are (see equation \ref{gspin}) :
\bq
\Sigma_*.G._\perp&=&+iG._\perp \label{confop3} \\
\Sigma_*.G_{*\perp}&=&-iG_{*\perp} \\
\Sigma_*.G_*.=\Sigma_*.G_{\perp\perp}&=& 0 \label{confop5}
\eq
where $\perp$ is a component in the plane orthogonal to the plane of the light-like vectors $z_\mu$ and $p_\mu$ (the projection operator associated to this plane is $g^\perp_{\mu\nu}$, defined by the equation \ref{gperp}).

If one has a field $\Phi(uz)$ ($u$ is a real number) which has a fixed projection $s$ of the Lorentz spin on to the line $z_\mu$, the generators of the collinear conformal subgroup act in the following way :
\bq
[J_+,\Phi(uz)]&=&\frac{1}{\sqrt{2}}\frac{d}{du}\Phi(uz) \label{diffconf1} \\
{}[J_3,\Phi(uz)]&=&\frac{1}{2}\left(l+s+2u\frac{d}{du}\right)\Phi(uz) \\
{}[J_-,\Phi(uz)]&=&\sqrt{2}\left(u(l+s)+u^2\frac{d}{du}\right)\Phi(uz) \\
{}[E,\Phi(uz)]&=&\frac{1}{2}(l-s)\Phi(uz). \label{diffconf4}
\eq

From this relation the irreducible representations of the collinear conformal subgroup $SO(2,1)$ which contains the non-trivial conformal
transformation on operators on the light-cone can be constructed. These representations are classified by the eigenvalues of the Casimir operator $J^2$. Equations \ref{confop1} to \ref{confop5} describe the construction of operators $\Phi(uz)$ which are eigenvectors of $J^2$ :

\be
[J^2,\Phi(uz)]=j(j-1)\Phi(uz)
\ee
where 
\be
j=\frac{1}{2}(l+s) \label{confspin}
\ee
$j$ is called "conformal spin". For $\Phi(uz)$, the algebra is reduced to differential operators, so one can work with functions of one real
variable instead of quantum fields. Having in mind the application to the computation of wave functions, one can make the Fourier Transform of the quantum field $\Phi$ :

\be
\Phi(uz)=\int d\alpha e^{-iu\alpha pz}\phi(\alpha).
\ee

In this formalism, the irreducible representations of the collinear conformal subgroup can be built, classified by the eigenvalues of $J^2$ and $-J_3$. They are the set of the following functions :
\be
|j,n\rangle =\frac{1}{\Gamma(j+n)}\left(\frac{\alpha}{i\sqrt{2}}\right)^{j+n-1}
\label{irrep}
\ee

Equations \ref{diffconf1} to \ref{diffconf4} give the action of the generators of the collinear conformal subgroup on the states $|j,n\rangle$ :

\bq
J_+|j,n\rangle&=&(j+n)|j,n+1\rangle \\
{}J_-|j,n\rangle&=&(n-j)|j,n-1\rangle \\
{}J_3|j,n\rangle&=&-n|j,n\rangle.
\eq

One can see that $J_+$ transforms a state one step upper and $J_-$ one step lower. The lowest value of $n$ is $j$. 

To build a local conformal operator on the light-cone, one has to have fixed projection of the Lorentz spin on to the light-cone (equations \ref{confop1} to \ref{confop5}). If the operator is not taken at $z=0$, one has to expand it in the
states \ref{irrep}. But in order to compute wave-functions, the $z$-dependence of polylocal operators must be controlled, like those in \ref{2matel} and \ref{3matel}.

%

\subsubsection{Polylocal operators on the light-cone}

Suppose that $\Phi_1(u_1z),\ldots,\Phi_n(u_nz)$ are $n$ local operators which have a fixed conformal spin. What happens for the product $\Phi_1(u_1z)\ldots\Phi_n(u_nz)$ ? When one thinks of irreducible representation of the collinear conformal subgroup, the corresponding "Clebsch-Gordon coefficients" of a tensor product of different irreducible representations are needed :
\be
|j,n\rangle =\sum_{n_1+\ldots+n_k=n}C^{j,n}_{j_1,n_1,\ldots,j_k,n_k}
|j_1,n_1\rangle\ldots|j_k,n_k\rangle \label{CGcoeff}
\ee 

This seems a very complicated task, but one has to recall what is the conformal collinear subgroup : it is the Lorentz group in $2+1$ dimensions ($SO(2,1)$). So the irreducible representations $|j,n\rangle$ can be interpreted as relativistic particles in an abstract or internal space of $2+1$ dimensions. With this point of view, the state $|j,n\rangle$ is a particle with a mass $j$ and an energy $n$ (that is why $n\ge j$). 

Consider a system of $k$ particles. The lowest invariant mass of
this system is the sum of the masses of all particles, $j=j_1+\ldots+j_k$. If the state of the lowest invariant mass has also the lowest energy, all the particles must be a their lowest energy level, $n_i=j_i$. This state is non-degenerate. In the language of irreducible representations of $SO(2,1)$, it gives~:
\bq
\left|j_{\text{min}}=\sum_{i=1}^{k}j_i,n_{\text{min}}=j_{\text{min}}\right
\rangle & \sim &
|j_1,j_1\rangle\ldots|j_k,j_k\rangle \\
&\sim&\alpha_1^{2j_1-1}\ldots\alpha_k^{2j_k-1}
\eq 

If $n>j$, the operator $J_+=\sum_{i=1}^{k}J_+^{(i)}$ can be used to "raise" the "energy" of the state. Since $J_+\phi(\alpha)=\frac{1}{\sqrt{2i}}\alpha\phi(\alpha)$, one gets :
\bq
\left|j_{\text{min}}=\sum_{i=1}^{k}j_i,n\right\rangle &\sim&
(\alpha_1+\ldots+\alpha_k)^{n-j}|j_1,j_1\rangle\ldots|j_k,j_k\rangle \\
&\sim&(\alpha_1+\ldots+\alpha_k)^{n-j}\alpha_1^{2j_1-1}\ldots\alpha_k^{2j_k-1}
\label{jminCG}
\eq 

To have the whole sum for the state $|j,n\rangle$ (equation \ref{CGcoeff}), not only the states at rests $|j_i,j_i\rangle$ are needed but also the higher ones ($|j_i,n_i\rangle$ with $n_i>j_i$). In that case, there are different possibilities and the Clebsch-Gordan coefficients contains binomial coefficients (see \cite{Oh}). For a bilocal operator, one gets :
\bq
|j,n\rangle= (\alpha_1+\alpha_2)^{n-j}\sum_{n_1+n_2=j-j_1-j_2}\left(
	\begin{array}{c} 
	n_1 +n_2 \\ n_1
	\end{array}
\right)|j_1,j_1+n_1>|j_2,j_2+n_2> \nonumber \\ \label{bilocCG}
\eq

The summation gives :

\be
|j,n>\sim (\alpha_1+\alpha_2)^{n+j+1}(1+\xi)^{2j_1-1}(1-\xi)^{2j_2-1}
P_{j-j_1-j_2}^{(2j_1-1,2j_2-1)}(\xi) \label{repJa}
\ee
with
\be
\xi=\frac{\alpha_1-\alpha_2}{\alpha_1+\alpha_2}
\ee
where $P_n^{(\nu_1,\nu_2)}(\xi)$ are the Jacobi polynomials\cite{Er} (see appendix \ref{app:a}). Analogously, for product of three local operators, one gets Appell polynomials\cite{Er} (see appendix \ref{app:a}). The important thing to remark about equation \ref{repJa} is that the only dependence in $n$ is in the factor $(\alpha_1+\alpha_2)^{n+j+1}$, the other parts of this equation depend only of the conformal spins $j_1$, $j_2$ and $j$.

An important property will be needed for the next subsection. For any kind of irreducible representation of the collinear conformal subgroup $|j, n\rangle$ which is constructed from the tensor product of $k$-irreducible representations, there is the following homogeneity property :
\be
|j,n\rangle \equiv \phi_{j,n}(\alpha_1,\ldots ,\alpha_k)=(\alpha_1+\ldots+\alpha_k
)^{n+j-1}\phi_{j,n}(\hat{\alpha}_1,\ldots,\hat{\alpha}_k) \label{hom}
\ee
where
\be
\hat{\alpha}_i=\frac{\alpha_i}{\alpha_1+\ldots+\alpha_k}
\ee

This property comes almost directly from the constructions of the Clebsch-Gordon
coefficients (equations \ref{jminCG} and \ref{bilocCG}). Hence irreducible representations $|j,n\rangle$ of the collinear conformal subgroup on the space of $k$ variables induce the irreducible representations on the functions defined on the simplex $\alpha_1+\ldots+\alpha_k=1$. This property will be used for the next subsection.

%

\subsection{Matrix element of polylocal conformal operators}

Suppose that one has the matrix element of a product of local conformal operators $\Phi_1(u_1z)\ldots\Phi_k(u_k)$ on the light-cone between the vacuum and a massless meson state $h$ of momentum $p_\mu$. This matrix element can be parameterized like this :
\be
\langle 0|\Phi_1(u_1z)\ldots\Phi_k(u_kz)|h(p)\rangle=\int {\cal D}\alpha_i
e^{-ipz(u_1\alpha_1+\ldots+u_k\alpha_k)}\phi(\alpha_1,\ldots,\alpha_k)
\ee
where
\be
\int{\cal D}\alpha_i=\int d\alpha_1\ldots d\alpha_k\delta\left(\sum_{i=1}^{k}
\alpha_i-1\right)
\ee

According to the property of equation \ref{hom}, the wave function $\phi(\alpha_1,\ldots,\alpha_k)$ can be expanded in different parts which have a fixed conformal spin $j$. The minimum value of the conformal spin is the sum of the conformal spins of every conformal operators $\Phi_i(u_iz)$. The different parts of this conformal expansion are mutually orthogonal polynomials (Jacobi polynomials if there are only two operators) and form the complete set of functions on the simplex $\alpha_1+\ldots+\alpha_k=1$.

In front of each of these polynomials there is a coefficient which is renormalized multiplicatively. This conformal expansion is justified by the construction of irreducible representations of the collinear conformal subgroup.

The first term of such an expansion is called "asymptotic" wave function and
can be easily computed (equation \ref{jminCG}) :
\be
\phi_{\text{as}}(\alpha_1,\ldots,\alpha_k)=
\frac{\Gamma(2j_1+\ldots+2j_k)}{\Gamma(2j_1)\ldots\Gamma(2j_k)}\alpha_1^{2j_1-1}
\ldots\alpha_k^{2j_k-1} \label{as}
\ee
with
\be
\int{\cal D}\alpha\phi_{\text{as}}(\alpha_i)=1
\ee

Now one can understand better the meaning of the example given at the beginning of this section :
\be
\phi_\| (u)=6u\ubar \sum^{\infty}_{n=0}a^\|_nC_n^{3/2} (\xi). \label{confex2}
\ee

$\phi_\| (u)$ is the contribution of the matrix element $\langle 0|\ov{u} (z) \gamma_\mu [z,-z] d ( -z)|\rho^- (P, \lambda) \rangle$ where each quark field has a positive spin projection $s=+\frac{1}{2}$ (see equation \ref{confop1}). Knowing that a quark field has the canonical dimension $l=\frac{3}{2}$, each quark field has a conformal spin $j_q=1$ (see equation \ref{confspin}). So the asymptotic distribution amplitude \ref{as} equals $\phi_{\text{as}}(\alpha_q, \alpha_{\ov{q}})=6\alpha_q\alpha_{\ov{q}}$ and has the conformal spin $j=2$. The higher terms in \ref{confex2} corresponds of higher values of $j$ and $n$ in equation \ref{repJa}. Denoting $u=\alpha_q$ with $\alpha_q+\alpha_{\ov{q}}=1$, one gets the expansion \ref{confex2} (the Gegenbauer polynomials $C_n^{3/2}$ are proportional to  the Jacobi ones $P^{(1,1)}_n$ (see Appendix C) which appear when $j_1=j_2=1$, see \ref{repJa}).

%
%

\sect{Example of computation \label{PsiPhi}}

In this section, the method of computing wave functions (defined in the appendix A) is illustrated with non-leading twist three-point distributions for the $\rho$-meson (following \cite{BBS}).

One starts with the definition of the chiral-even three-points distributions~:

\bq
\lefteqn{\langle 0|\bar u(z) i\gamma_\alpha [z,vz]gG_{\mu\nu}(vz)[vz,-z] 
	d(-z)|\rho^-(P,\lambda)\rangle=}\makebox[2cm]{\ }\nonumber \\
&=&   f_{\rho}m_\rho\left[ p_\alpha[p_\mu e^{(\lambda)}_{\perp\nu}-p_\nu
	e^{(\lambda)}_{\perp\mu}]{\cal V}(v,pz)\right. \nonumber \\
& &{}+m_\rho^2\frac{\epl\cdot z}{p\cdot z}[p_\mu g_{\alpha\nu}^\perp
	-p_\nu g_{\alpha\mu}^\perp]\Phi(v,pz) \nonumber \\
& &{}\left.+m_\rho^2\frac{\epl\cdot z}{(p\cdot z)^2}[p_\mu z_\nu-p_\nu z_\mu]
	p_\alpha \Psi(v,pz) \right] \nonumber \\
& &{}+\text{higher twists}\label{exechie3}
\eq
\bq
\lefteqn{\langle 0|\bar u(z) \gamma_\alpha \gamma_5
        [z,vz]g\Gtilde_{\mu\nu}(vz)[vz,-z] 
	d(-z)|\rho^-(P,\lambda)\rangle=}\makebox[2cm]{\ }\nonumber \\
&=&if_{\rho}m_\rho\left[ p_\alpha[p_\mu e^{(\lambda)}_{\perp\nu}-p_\nu        	
	e^{(\lambda)}_{\perp\mu}]{\cal A}(v,pz)\right. \nonumber \\
& &{}+m_\rho^2\frac{\epl\cdot z}{p\cdot z}[p_\mu g_{\alpha\nu}^\perp
	-p_\nu g_{\alpha\mu}^\perp]\tilde{\Phi}(v,pz) \nonumber \\
& &{}\left.+m_\rho^2\frac{\epl\cdot z}{(p\cdot z)^2}[p_\mu z_\nu-p_\nu z_\mu]
	p_\alpha \tilde{\Psi}(v,pz) \right] \nonumber \\
& &{}+\text{higher twists} \label{exechie3b}
\eq
 
One wants to compute the non-leading twist distributions $\Phi(v,pz)$, $\Psi(v,pz)$, $\tilde{\Phi}(v,pz)$ and $\tilde{\Psi}(v,pz)$. The leading twist distributions have the (following) conformal expansion (see \cite{BBKT}) :
\be
A(v,pz)=\int{\cal D}\un \alpha e^{ipz(\alpha_u-\alpha_d+v\alpha_g)}A(\un \alpha)
\ee
\be
A(\un \alpha)=360 \zeta_3\alpha_d\alpha_u\alpha_g^2
\left[1+w_3^A\frac{1}{2}(7\alpha_g-3)\right]+\ldots
\ee
\be
V(v,pz)=\int{\cal D}\un \alpha e^{ipz(\alpha_u-\alpha_d+v\alpha_g)}V(\un \alpha)
\ee
\be
V(\un \alpha)=\zeta_3\omega_3^V[(\alpha_d-\alpha_u)\alpha_d\alpha_u\alpha_g^2
+\ldots]
\ee

To compute $\Phi(v,pz)$ and $\tilde \Phi(v,pz)$, the wave function of conformal operators is needed~:
\be
\langle 0|\bar u(z)g\tilde G_{\mu\nu}\gamma.\gamma_\alpha\gamma_5\gamma_*d(-z)
|\rho(P,\lambda)\rangle=f_\rho m_\rho^3\frac{\epl\cdot z}{p\cdot z}\left[
p_\mu g_{\alpha\nu}^\perp-p_\nu g_{\alpha\mu}^\perp\right]\Phi^{\ud}(v,pz)
\label{t3conf1}
\ee
\be
\langle 0|\bar u(z)g\tilde G_{\mu\nu}\gamma_*\gamma_\alpha\gamma_5\gamma.d(-z)
|\rho(P,\lambda)\rangle=f_\rho m_\rho^3\frac{\epl\cdot z}{p\cdot z}\left[
p_\mu g_{\alpha\nu}^\perp-p_\nu g_{\alpha\mu}^\perp\right]\Phi^{\du}(v,pz)
\label{t3conf2}
\ee
(this parameterization corresponds to the highest conformal spin for $\tilde G_{\mu\nu}$, although this gluon field does not look explicit like a conformal operator, see equations \ref{confop3} to \ref{confop5}).

These two matrix elements have fixed conformal spin, so a conformal expansion (with the help of the Appell polynomials) can be done :
\be
\Phi^{\ud}(\un \alpha)=K^{\ud} 60\alpha_u\alpha_g^2\left[1+\omega_{1,0}^{\ud}
(\alpha_g-3/2\alpha_u)+\omega_{0,1}^{\ud}(\alpha_g-3\alpha_d)+\ldots\right]
\ee
\be
\Phi^{\du}(\un \alpha)=K^{\du} 60\alpha_d\alpha_g^2\left[1+\omega_{1,0}^{\du}
(\alpha_g-3\alpha_u)+\omega_{0,1}^{\du}(\alpha_g-3/2\alpha_d)+\ldots\right]
\ee

Then it is easy to get $\Phi$ and $\tilde \Phi$ from $\Phi^{\ud}$ and $\Phi^{\du}$ :
\bq
\tilde \Phi(\un \alpha)&=&\frac{1}{2}\left[\Phi^{\ud}(\un \alpha)+\Phi^{\du}
(\un \alpha)\right] \\
\Phi(\un \alpha)&=&\frac{1}{2}\left[\Phi^{\ud}(\un \alpha)-\Phi^{\du}
(\un \alpha)\right] 
\eq

The G-parity (charge conjugation $\times$ isospin) applied to the matrix elements \ref{t3conf1} and \ref{t3conf2} implies the following relations :
\bq
K^{\ud}&=&K^{\du} \\
\omega_{1,0}^{\ud}&=&\omega_{0,1}^{\du} \\
\omega_{1,0}^{\du}&=&\omega_{0,1}^{\ud}
\eq

Finally, the expression for $\Phi$ and $\tilde \Phi$ (knowing that $1-\alpha_g-\alpha_u-\alpha_d=0$) is the following~:
\bq
\tilde \Phi(\un \alpha)=K^{\ud}30\alpha_g^2\Big\{(1-\alpha_g)&+&
\omega_{1,0}^{\ud}\left[\alpha_g(1-\alpha_g)-3/2(\alpha^2_u-\alpha_d^2)\right]
 \nonumber \\
&+&\left.\omega_{0,1}^{\ud}\left[\alpha_g(1-\alpha_g)-
6\alpha_u\alpha_d\right]\right\}
\eq
\be
\Phi(\un \alpha)=K^{\ud}30\alpha^2_g(\alpha_u-\alpha_d)\left[1+
1/2\omega_{1,0}^{\ud}(5\alpha_g-3)+\omega_{0,1}^{\ud}\alpha_g\right]
\ee

Now, the different constants $K^{\ud}$, $\omega_{1,0}^{\ud}$ and $\omega_{0,1}^{\ud}$ has to be extracted from matrix elements of local operators.

To find $K^{\ud}$, the definition of $\tilde \Phi(v,pz)$ in \ref{exechie3b} is taken. It implies :
\be
f_\rho m_\rho^3(\epl\cdot z)g_{\mu\nu}\int{\cal D}\un \alpha \tilde 
\phi(\un \alpha)=\langle 0|\bar u\tilde G._{\mu}\gamma_\nu\gamma_5
d|\rho(P,\lambda)\rangle
\ee

Using the equation \ref{defxi} which defines $\zeta_3$ and $\zeta_4$,
\begin{eqnarray}
\langle0|\bar u g\tilde G_{\mu\nu}\gamma_\alpha
 \gamma_5 d|\rho^-(P,\lambda)\rangle  &=&
f_\rho m_\rho \zeta_{3}
\Bigg[
e^{(\lambda)}_\mu\Big(P_\alpha P_\nu-\frac{1}{3}m^2_\rho \,g_{\alpha\nu}\Big)  
-e^{(\lambda)}_\nu\Big(P_\alpha P_\mu-\frac{1}{3}m^2_\rho \,g_{\alpha\mu}\Big)
\Bigg]
\nonumber\\
&&{}+\frac{1}{3}f_\rho m_\rho^3 \zeta_{4}
\Bigg[e^{(\lambda)}_\mu g_{\alpha\nu}- e^{(\lambda)}_\nu g_{\alpha\mu}\Bigg]  
\label{exedefxi}
\end{eqnarray}
one obtains for $K^{\ud}$ :
\be
K^{\ud}=\left[-\frac{1}{3}\zeta_3+\frac{1}{3}\zeta_4\right] \label{Kud}
\ee

The two other constants $\omega_{1,0}^{\ud}$ and $\omega_{0,1}^{\ud}$ have to be computed. The idea is to relate these coefficients to local matrix elements. To do it, the matrix elements \ref{t3conf1} are expanded to the order $(z_\mu)$. It gives
\bq
\lefteqn{\langle 0|\bar ug\left(-\Dl .g\tilde G_{\cdot\nu}+g\tilde G_{\cdot\nu}\Dr .\right)
\gamma.\gamma_\alpha\gamma_5\gamma_*d|0\rangle=}\nonumber \\
&=&i\f \m^3(\epl\cdot z)(p \cdot z)g^\perp_{\alpha\nu}\int{\cal D}\un\alpha
(\alpha_u-\alpha_d)\Phi^{\ud}(\un\alpha) \label{derconft3a}
\eq
\be
\int{\cal D}\un\alpha
(\alpha_u-\alpha_d)\Phi^{\ud}(\un\alpha)=K^{\ud}\left(\frac{1}{6}+\frac{1}{14}
\omega_{0,1}^{\ud}\frac{1}{14}-\omega_{1,0}^{\ud}\right)
\ee
\bq
\langle 0|\bar ug\left(\Dl .g\tilde G_{\cdot\nu}\right)
\gamma.\gamma_\alpha\gamma_5\gamma_*d|0\rangle
&=&-i\f \m^3(\epl\cdot z)(p \cdot z)g^\perp_{\alpha\nu}\int{\cal D}\un\alpha
\alpha_g\Phi^{\ud}(\un\alpha) \nonumber \\ \\
\int{\cal D}\un\alpha
\alpha_g\Phi^{\ud}(\un\alpha)&=&K^{\ud}\left(\frac{1}{2}+\frac{1}{14}
\omega_{0,1}^{\ud}+\frac{1}{14}\omega_{1,0}^{\ud}\right). \label{derconft3d}
\eq

These equations can be also rewritten like this (replacing $K^{\ud}$ by its value in \ref{Kud}) :
\bq
\lefteqn{\langle 0 |\bar u\left(-i\Dl.gG._\nu+gG._\nu i\Dr.\right)
i\gamma_\alpha d|\rho(P,\lambda)\rangle =}\nonumber \\
&=&-\f\m^3\epl.P.g_{\alpha\nu}\left[\frac{1}{6}\left(-\frac{1}{3}\zeta_3 +\frac{1}{3}\zeta_4\right)+\frac{1}{14}R_1-\frac{1}{14}R_2\right] +\ldots
\nonumber \label{Ra}\\
\eq
\bq
\lefteqn{\langle 0 |\bar u\left(i\Dr.g\tilde G._\nu\right)
\gamma_\alpha\gamma_5 d|\rho(P,\lambda)\rangle =}\nonumber \\
&=&\f\m^3\epl.P.g_{\alpha\nu}\left[\frac{1}{2}\left(-\frac{1}{3}\zeta_3 +\frac{1}{3}\zeta_4\right)+\frac{1}{14}R_1+\frac{1}{14}R_2\right] +\ldots
\nonumber \label{Rb}\\
\eq
where
\bq
R_1\equiv K^{\ud}\omega_{0,1}^{\ud} \label{defR1} \\
R_2\equiv K^{\ud}\omega_{1,0}^{\ud} \label{defR2}
\eq

Because of the G-parity symmetry, one can suspect that only two local operators are involved (apart from operators with total derivative). One defines :
\bq
O^1_{\alpha\beta\mu\nu}&=&\bar u\left(i\Dr_\beta g\tilde G_{\mu\nu}+g\tilde G_{\mu\nu}
i\Dr_\beta\right)\gamma_\alpha\gamma_5d \\
O^2_{\alpha\beta\mu\nu}&=&\bar u\left(-\Dl_\beta g G_{\mu\nu}+g G_{\mu\nu}
\Dl_\beta\right)\gamma_\alpha d
\eq

The different Lorentz-structures for $O^1_{\alpha\beta\mu\nu}$ are the following :
\bq
\lefteqn{\langle 0|O^1_{\alpha\beta\mu\nu}|\rho(P,\lambda)\rangle=} \nonumber \\
&=&\left\{\epl_\mu\left[P_\alpha P_\beta P_\nu-\frac{5}{24}\m^2\left(P_\alpha
g_{\beta\nu}+P_\beta g_{\alpha\nu}\right)-\frac{1}{6}\m^2P_\nu g_{\alpha\beta}
\right]\right. \nonumber \\
& &{}-\epl_\nu\left[P_\alpha P_\beta P_\nu-\frac{5}{24}\m^2\left(P_\alpha
g_{\beta\mu}+P_\beta g_{\alpha\mu}\right)-\frac{1}{6}\m^2P_\mu g_{\alpha\beta}
\right] \nonumber \\
& &{}\left.-\frac{1}{24}\m^2\left[\epl_\alpha\left(g_{\beta\nu}P_\mu
-g_{\beta\mu}P_\nu\right)+\epl_\beta\left(g_{\alpha\nu}P_\mu-g_{\alpha\mu}P_\nu
\right)\right]\right\}A_1\f\m \nonumber \\
& &{}+P_\alpha\left(\epl_\mu g_{\beta\nu}-\epl_\nu g_{\beta\mu}\right)B_1
+P_\beta\left(\epl_\mu g_{\alpha\nu}-\epl_\nu g_{\alpha\mu}\right)C_1 \nonumber \\
& &{}+\epl_\alpha\left(P_\mu g_{\beta\nu}-P_\nu g_{\beta\mu}\right)D_1 +
\epl_\beta\left(P_\mu g_{\alpha\nu}-P_\nu g_{\alpha\mu}\right)E_1 \nonumber \\
& &{}-g_{\alpha\beta}\left(P_\mu\epl_\nu-P_\nu\epl_\mu\right) \label{defO1}
\eq
And the constants $A_2$ to $F_2$ can be similarly defined, replacing $O^1$ by $O^2$ in this last equation.

Traces can be taken in order to reduce the number of independent constants (it works for both $O^1$ and $O^2$) :
\be
g_{\alpha\beta} : \;B+C-D-E+4F=0
\ee
\be
g_{\alpha\mu} : \;-\epl_\nu P_\beta(B+3C+F)-\epl_\beta P_\nu(D+3E-F)=
\langle 0|O_{\xi\beta\xi\nu}|\rho(P,\lambda)\rangle
\ee
\be
g_{\beta\mu} : \;-\epl_\alpha P_\nu(3B+C+F)-\epl_\nu P_\alpha(3D+E-F)=
\langle 0|O_{\alpha\xi\xi\nu}|\rho(P,\lambda)\rangle
\ee

One defines $a^1$, $a^2$, $b^1$ and $b^2$ in the following way :
\be
\langle 0|O^{1,2}_{\xi\beta\xi\nu}|\rho(P,\lambda)\rangle
=(\epl_\beta P_\nu+\epl_\nu P_\beta)a_{1,2} \f\m^3 + \text{higher twist}
\label{defa1}
\ee
\be
\langle 0|O^{1,2}_{\alpha\xi\xi\nu}|\rho(P,\lambda)\rangle
=(\epl_\alpha P_\nu+\epl_\nu P_\alpha)b_{1,2} \f\m^3 + \text{higher twist}
\ee

These higher (theoretical-)twist terms are not taken into account, because they are bigger than twist 4. This approximation gives :
\bq
B&=&D=\frac{1}{8}(a-3b)\m^3\f \\
E&=&C=\frac{1}{8}(b-3a)\m^3\f \label{CE}\\
F&=&0
\eq

The goal is to express the unknown coefficients $\omega^{\ud}_{0,1}$ and $\omega_{1,0}^{\ud}$ in function of $A\cdots F$ and $a$, $b$.

For it, the definition of $\zeta_3$ and $\zeta_4$ in \cite{BBKT} (equation \ref{defxi}) is used. It can be written like this :
\be
\langle 0|\bar ug\tilde G._\nu\gamma_\alpha\gamma_5 d|\rho(P,\lambda)\rangle
=\epl.g_{\alpha\nu}\f\m^3\left(-\frac{1}{3}\zeta_3 +\frac{1}{3}\zeta_4\right) +
\ldots \label{xi34}
\ee
Equation \ref{defO1} implies :
\be
\langle 0|O^{1,2}_{\alpha\cdot\cdot\nu}|\rho(P,\lambda)\rangle =
e^\lambda _{\cdot} P.g_{\alpha\nu}\left(-\frac{1}{4}\f\m^3A_{1,2}+C_{1,2}+E_{1,2}\right)
+\ldots\label{O..}
\ee

Taking \ref{Ra} and \ref{Rb} with the total derivative of \ref{xi34} (with $i\partial .\rightarrow P.$), combining \ref{CE} with \ref{O..}, one gets :
\be
\frac{1}{6}\left(-\frac{1}{3}\zeta_3+\frac{1}{3}\zeta_4\right)+\frac{1}{14}R_1
-\frac{1}{14}R_2=-\frac{1}{4}[A_2-b_2+3a_2]
\ee
\be
\frac{1}{2}\left(-\frac{1}{3}\zeta_3+\frac{1}{3}\zeta_4\right)+\frac{1}{14}R_1
+\frac{1}{14}R_2=\left(-\frac{1}{3}\zeta_3 +\frac{1}{3}\zeta_4\right) +\frac{1}{4}[A_1-b_1+3a_1]
\ee

The constants $A_{1,2}$ and $B_{1,2}$ can be extracted from the coefficients of the
twist 3 wave functions $A(\un \alpha)$ and $V(\un \alpha)$. For that, the equations \ref{exechie3} and \ref{exechie3b} are expanded at the order $(z_\mu)$ :
\be
\langle 0 |\bar u\left(-\Dl_{\cdot}\gamma.gG._\nu+gG._\nu \gamma.\Dr_{\cdot}\right)
 d|\rho(P,\lambda)\rangle =
\f\m\epl_{\perp\nu}p_{\cdot}^3\frac{3}{28}\zeta_3\omega_3^V
\ee
\be
\langle 0 |\bar u\gamma.\gamma_5\left(i\Dr_{\cdot}g\tilde G._\nu\right)
 d|\rho(P,\lambda)\rangle =
-\f\m\epl_{\perp\nu}p_{\cdot}^3\zeta_3\left(\frac{3}{7} +\frac{3}{28}\omega_3^A\right)
\ee

The definition of $A_{1,2}$ applied to these equations gives :
\bq
A_1&=&\zeta_3\left(\frac{3}{7} +\frac{3}{28}\omega_3^A\right) \\
A_2&=&-\frac{3}{28}\zeta_3\omega_3^V.
\eq

Now, the three coefficients ($K^{\ud}$, $R_1$ and $R_2$ for $\Phi(\un \alpha)$ and
$\tilde \Phi(\un \alpha)$ are expressed in functions of constants coming for the twist 3 wave functions and some unknown coefficients $a_{1,2}$, $b_{1,2}$. Their number can be reduced using exact operator relations (with the equation of motion)~:

The equation
\be 
D_\mu\tilde G_{\mu\nu}=0
\ee
implies (after taking the matrix element)
\be
b_1=-\frac{1}{3}\zeta_3-\frac{1}{6}\zeta_4
\ee

The equation
\be
O^2_{\xi\beta\xi\alpha}-O^2_{\beta\xi\xi\alpha}=O^1_{\xi\alpha\xi\beta}
-O^1_{\alpha\xi\xi\beta}+g_{\alpha\beta}\left[i\bar u\gamma_\sigma
\gamma_5\left(\tilde G_{\xi\sigma}\Dl_\xi+\Dr_\xi\tilde G_{\xi\sigma}\right)d
\right]
\ee
implies
\be
a_2-b_2=a_1-b_1
\ee

And the equation
\bq
\lefteqn{6\partial_\nu O_{\mu\alpha\beta} =}\nonumber \\
&=&-12i\bar u\gamma_\rho\left(G_{\rho\beta}\Dr_\alpha -\Dl_\alpha G_{\rho\beta}
+(\alpha \leftrightarrow)\right) d \nonumber \\
& &{}+\frac{1}{3}\partial^2\left(\partial_\alpha\bar u\gamma_\beta d+
\partial_\beta\bar u\gamma_\alpha d\right)-4\partial_\rho\bar u
\left(\gamma_\beta\tilde G_{\alpha\rho}+\gamma_\alpha\tilde G_{\beta\rho}\right)
\gamma_5 d \nonumber \\
& &{}-\frac{8}{3}\partial_\beta\bar u\gamma_\sigma\tilde G_{\sigma\alpha}
\gamma_5 d-\frac{8}{3}\partial_\alpha \bar u\gamma_\sigma\tilde G_{\sigma\beta}
\gamma_5 d \nonumber \\
& &{}+\frac{28}{3}g_{\alpha\beta}\partial_\rho\bar u\gamma_\sigma
\tilde G_{\sigma\rho} d
\eq
implies
\be
24a_2=-\frac{5}{3}\frac{1}{\f\m}\langle\langle O_2\rangle\rangle+
\frac{1}{3}\frac{1}{\f\m}\langle\langle O_0\rangle\rangle+\frac{8}{3}\zeta_3
-\frac{4}{3}\zeta_4
\ee
with $O_{\mu\alpha\beta}$ the following leading twist operator (symmetric and traceless) :
\be
O_{\mu\alpha\beta}=\bar u\left(\gamma_\mu D_\alpha D_\beta\right)_{\text{sym.}}
 d-\text{traces}
\ee

and $\langle\langle O_n\rangle\rangle$ the different moments of the leading twist part of the two-point wave function $\phi_\|$ :
\be 
\langle\langle O_n\rangle\rangle=M_n^{\phi_\|}
\ee
with
\be
M_n^{\phi_\|}=\int du \xi^n\phi_\|(u)
\ee
($\langle\langle O_n\rangle\rangle$ is also defined by equation \ref{defOn}).

Taking everything together, one gets :
\bq
\frac{1}{7}R_2&=&\frac{1}{4}(A_1+A_2)+a_1-\frac{5}{144}\frac{1}{\f\m}
\langle\langle O_2\rangle\rangle+\frac{1}{144}\frac{1}{\f\m}
\langle\langle O_0\rangle\rangle+\frac{5}{18}\zeta_4
\nonumber \\ \\
\frac{1}{7}R_1&=&\frac{1}{4}(A_1-A_2)+\frac{1}{2}a_1+\frac{5}{144}\frac{1}{\f\m}
\langle\langle O_2\rangle\rangle-\frac{1}{144}\frac{1}{\f\m}
\langle\langle O_0\rangle\rangle-\frac{1}{6}\zeta_3+\frac{5}{36}\zeta_4
\nonumber \\
\eq

So the unknown coefficients $R_{1,2}$ (equation \ref{defR1} and \ref{defR2}) which are in the wave functions $\Phi$ and $\tilde \Phi$ are expressed in function of known coefficients, expect for $a_1$ which has to be computed by the method of QCD sum rules.

With the same method, $\Psi$ and $\tilde \Psi$ can be computed :
\bq
\Psi(\un \alpha)&=&N_\Psi 1260(\alpha_d\alpha_u)\alpha_d\alpha_u\alpha_g \\
\tilde \Psi(\un\alpha)&=&120\alpha_u\alpha_d\alpha_g\left[\tilde\Psi_0
+\tilde \Psi_1(3\alpha_g-1)\right]
\eq
with
\bq
N_\Psi&=&-\frac{5}{72}\langle\langle O_2\rangle\rangle+\frac{1}{72}\langle\langle
O_0\rangle\rangle-\frac{1}{2}a_1-\frac{1}{2}A_2-\frac{1}{18}\zeta_3
-\frac{5}{36}\zeta_4 \nonumber \\ \\
\tilde \Psi_0&=&\frac{2}{3}\zeta_3+\frac{1}{3}\zeta_4 \\
\frac{2}{21}\tilde \Psi_1&=&\frac{1}{2}a_1-\frac{1}{2}A_1
+\frac{5}{18}\zeta_3+\frac{5}{36}\zeta_4
\eq

So, in order to have the non-leading twist wave functions $\Phi$, $\tilde \Phi$, $\Psi$ and
$\tilde \Psi$, the coefficient $a_1$ has to be computed. But it is defined from the matrix element of a local operator in equation \ref{defa1}, hence the method of two-point QCD sum rule described in \cite{SVZ} can be used.

In \cite{BBKT}, there is a good presentation of the renormalization dependence of the coefficients appearing in the wave functions. The basic idea is to defined them from local conformal operators, where the renormalization dependence is better known (see section \ref{sec:confQCD}).

%
%

\sect{Meson mass correction}

In this section, some new techniques (partially reproduced in \cite{BBS}) to compute mass corrections of wave functions are described.

The mass of the $\rho$-meson induce some non-leading (physical-)twist structure
for the non-local matrix element. 

For example, for the chiral-even two-point distributions
\begin{eqnarray}
\lefteqn{\langle 0|\bar u(x) \gamma_{\mu} [x,-x] 
d(-x)|\rho^-(P,\lambda)\rangle = }\makebox[2cm]{\ } \nonumber \\
&=& f_{\rho} m_{\rho} \left[ P_{\mu}
\frac{e^{(\lambda)}\cdot x}{P \cdot x}
\int_{0}^{1} \!du\, e^{i \xi P \cdot x} \left(A^{(e)}(u)+
\frac{x^2 m_\rho^2}{4}\hat{A}^{(e)}(u)\right) \right. 
\nonumber \\
& &{}+ \left(e^{(\lambda)}_{\mu}-P_\mu\frac{\epl\cdot x}{P\cdot x}\right)
\int_{0}^{1} \!du\, e^{i \xi P \cdot x} \left(B^{(e)}(u) 
+\frac{x^2 m_\rho^2}{4}\hat{B}^{(e)}(u)\right)\nonumber \\
& & \left.{}- \frac{1}{2}x_{\mu}
\frac{e^{(\lambda)}\cdot x }{(P \cdot x)^{2}} m_{\rho}^{2}
\int_{0}^{1} \!du\, e^{i \xi P \cdot x} C(u)^{(e)} \right]
\end{eqnarray}
some wave functions  would
disappear if the mass of the $\rho$ were $0$. Hence mass corrections are higher
(physical-)twist contributions to non-local matrix elements (in the previous example,  $\hat{A}^{(e)}(u)$ and $C^{(e)}(u)$ are twist 4, $\hat{B}^{(e)}$ is twist 5). In fact, the $m^2_\rho$ contribution can first be seen in the difference between the light-like basis ($p_\mu$, $z_\mu$ and $\epl_{\perp\mu}$ defined in \ref{defp}, \ref{defz} and \ref{defeperp}) needed for the conformal expansion and the vectors which characterize the $\rho$-meson ($P_\mu$ and $\epl_\mu$). These different basis introduce $m^2_\rho$ terms
which are sort of "explicit" and "direct" corrections; but there are also the
other corrections like $\hat{A}^{(e)}(u)$, $\hat{B}^{(e)}$ and $C^{(e)}(u)$.

%

\subsection{Example of mass correction for a scalar theory \label{sbscalar}}

Suppose that $\phi(x)$ is a scalar field with $M$ a scalar particle of momentum $P_\mu$ and mass $m$. If $x^2$ is almost on the light-cone, we have the following expansion :
\be
\langle 0|\phi(x)\phi(-x)|M(P)\rangle=\int_0^1du e^{i\xi Px}\left[g(u)+
\frac{1}{4}m^2x^2\hat{g}(u)+\ldots \right] \label{scanloc}
\ee
with
\be
\xi=2u-1
\ee

It is quite hard to compute directly $\hat{g}(u)$. In fact, the following approximation is done : in the Operator Product Expansion of $\phi(x)\phi(-x)$, only the leading-twist local operators (it is the "theoretical twist", see section \ref{deftw}) is taken. The leading (theoretical-)twist expansion of \ref{scanloc} is defined in the following way :  
\be
\langle 0|\left[\phi(x)\phi(-x)\right]_{\text{l. t.}}|M(P)\rangle=\int_0^1du e^{i\xi Px}\left[f(u)+
\frac{1}{4}m^2x^2\hat{f}(u)+\ldots \right] \label{ltsca}
\ee

In that case, $\hat{f}(u)$ can be computed in term of $f(u)$. The first way to do it is the "brute force" method, i. e. doing a local expansion of \ref{ltsca} :
\be
\langle0|\left[\phi\left(-\Drl x\right)^n\phi\right]_{\text{l. t.}}
|m\rangle=(iPx)^nM_n+\frac{1}{4}m^2x^2n(n-1)\hat{M}_{n-1}(iPx)^{n-2} \label{scmom}
\ee
where
\bq
M_n&=&\int_0^1du\xi^nf(u) \nonumber \\
\hat{M}_n&=&\int_0^1du\xi^n\hat{f}(u) \label{defmom}
\eq

The leading twist operator is always symmetrical in all his indices and traceless.

It implies :
\bq
\lefteqn{i^n\langle0|\left[\phi \Drl_{\nu_1}\ldots  \Drl_{\nu_n}\phi\right]
_{\text{l. t.}}|M(P)\rangle=\langle\langle O_n \rangle\rangle\times} \nonumber \\
&\times&\left\{P_{\nu_1}\ldots P_{\nu_n}-\frac{m^2}{2n}
\sum_{i<j}^ng_{\nu_i\nu_j}P_{\nu_1}\ldots P_{\nu_{i-1}}P_{\nu_{i+1}}\ldots 
P_{\nu_{j-1}} P_{\nu_{j+1}} \ldots P_{\nu_n}+O(m^4)\right\} \nonumber \\
\eq

Hence
\bq
\lefteqn{\langle0|\left[\phi\left(i\Drl x\right)^n\phi\right]_{\text{l. t.}}
|m\rangle=} \nonumber \\
&=&\left\{(Px)^n-\frac{m^2x^2}{2n}\cdot\frac{n(n-1)}{2}(Px)^{n-2}+O(m^4)\right\}
\langle\langle O_n\rangle\rangle
\eq

The comparison of this last equation with \ref{scmom} gives :
\bq
M_n&=&\langle\langle O_n\rangle\rangle \\
n\hat{M}_{n-2}&=&\langle\langle O_n\rangle\rangle=M_n 
\eq

The integrated version of these equations is
\be
\hat{f}(u)=2(2u-1)\theta(u-1/2)\int_u^1dtf(t)+2(1-2u)\theta(1/2-u)\int_0^udtf(t).
\label{fchap}
\ee

The same answer can be obtained using the techniques developed in \cite{BaBr} directly in an integrated form :
\be
\langle 0|\left[\phi(x)\phi(-x)\right]_{\text{l. t.}}|M(P)\rangle=\int_0^1du \left[e^{i\xi Px}\right]_{\text{l. t.}}f(u)
\ee
where
\be
\left[e^{i\xi Px}\right]_{\text{l. t.}}f(u)=e^{i\xi Px}+\frac{m^2x^2}{4}\xi^2
\int_0^1dvve^{i\xi vPx}+O(m^4)
\ee

It can be checked that these last equations reproduce \ref{fchap} :
\bq
\lefteqn{\int_0^1duf(u)\xi^2\int_0^1dv v e^{iv\xi Px}=}\nonumber \\
&=&2\int_0^1 dt e^{iPx(2t-1)}\int_0^1duf(u)\xi^2\int_0^1dv v
\delta(2t-1-\xi v)
\eq
and
\bq
\int_0^1dv v\delta(2t-1-\xi v)&=&\frac{1}{|\xi |}\frac{2t-1}{2u-1}
\left[\theta(t-1/2)\theta(u-t)+\theta(-t+1/2)\theta(t-u)\right] \nonumber 
\\ \\
&=&\frac{1}{\xi^2}(2t-1)\left[\theta(t-1/2)\theta(u-t)-\theta(1/2-t)\theta(t-u)
\right] \nonumber \\
\eq

The goal is to reproduce this formalism in the physical case : a vector $\rho$-meson in QCD.

%

\subsection{Mass correction for two-point wave function : theoretical twist method \label{thtmasscorr}}

\subsubsection{Leading theoretical twist}

The idea is to compute the mass corrections in the leading theoretical twist approximation. In this part, the case with the two-point chiral even distributions is computed.

The different wave functions are :
\begin{eqnarray}
\lefteqn{\langle 0|\bar u(x) \gamma_{\mu} [x,-x] 
d(-x)|\rho^-(P,\lambda)\rangle = }\makebox[2cm]{\ } \nonumber \\
&=& f_{\rho} m_{\rho} \left[ P_{\mu}
\frac{e^{(\lambda)}\cdot x}{P \cdot x}
\int_{0}^{1} \!du\, e^{i \xi P \cdot x} \left(A^{(e)}(u)+
\frac{x^2 m_\rho^2}{4}\hat{A}^{(e)}(u)\right) \right. 
\nonumber \\
& &{}+ \left(e^{(\lambda)}_{\mu}-P_\mu\frac{\epl\cdot x}{P\cdot x}\right)
\int_{0}^{1} \!du\, e^{i \xi P \cdot x} \left(B^{(e)}(u) 
+\frac{x^2 m_\rho^2}{4}\hat{B}^{(e)}(u)\right)\nonumber \\
& & \left.{}- \frac{1}{2}x_{\mu}
\frac{e^{(\lambda)}\cdot x }{(P \cdot x)^{2}} m_{\rho}^{2}
\int_{0}^{1} \!du\, e^{i \xi P \cdot x} C^{(e)}(u) \right]
\label{tpmasscor}
\end{eqnarray}

The "brute force" method, i. e. the expansion of the distributions in moments ($x\rightarrow 0$) can be used. It gives :
\bq
\lefteqn{\langle 0|\ov{u}\gamma_\mu(i\Drl x)^n d|\rho^-(P,\lambda)\rangle
=} \nonumber \\
&=&P_\mu\frac{\epl\cdot x}{P\cdot x}\left[(Px)^nM_n^A-\frac{x^2m_\rho^2}{4}
n(n-1)\hat{M}^A_{n-2}(px)^{n-2}\right] \nonumber \\
& &{}+ \left(e^{(\lambda)}_{\mu}-P_\mu\frac{\epl\cdot x}{P\cdot x}\right)
\left[(Px)^nM_n^B-\frac{x^2m^2_\rho}{4}n(n-1)\hat{M}^B_{n-2}(Px)^{n-2}\right]
\nonumber \\
& &{}-\frac{1}{2}x_\mu\frac{\epl\cdot x}{(P\cdot x)^2}m_\rho^2(Px)^nM_n^c
\eq
 where the moments $M_n^A$, $M_n^B$, $M_n^C$ $\hat{M}_n^A$ and $\hat{M}_n^B$ are defined like in \ref{defmom}.

The leading (theoretical-)twist of the operators in the left hand side can be written. The condition of symmetry and zero trace gives :
\bq
\lefteqn{\langle 0|\ov{u}\gamma_\mu(iD^{\leftrightarrow}x)^n d|\rho^-(P,\lambda)\rangle=} \nonumber \\
&=&\left\{{} P_\mu\frac{\epl\cdot x}{P\cdot x}(Px)^n+\frac{1}{n+1}
\left(e^{(\lambda)}_{\mu}-P_\mu\frac{\epl\cdot x}{P\cdot x}\right)
\left[(Px)^n-\mx\frac{n(n-1)}{n+1}(Px)^{n-2}\right]\right. \nonumber \\
& &{}-\left. \mx P_\mu\frac{\epl\cdot x}{P\cdot x}(Px)^{n-2}
n\frac{(n-1)^2}{(n+1)^2}-\frac{1}{2}x_\mu\frac{\epl\cdot x}{(P\cdot x)^2}
(Px)^nm_\rho^2\frac{(n-1)n}{(n+1)^2}\right\}\langle\langle O_n\rangle\rangle
\nonumber \label{defOn} \\
\eq

The comparison of these two equations implies :
\bq
M_n^A&=&\langle\langle O_n\rangle\rangle \label{ltmom1} \\
M_n^B&=&\frac{1}{n+1}\langle\langle O_n\rangle\rangle \\
\hat{M}^A_{n-2}&=&\frac{n-1}{(n+1)^2}\langle\langle O_n\rangle\rangle \\
\hat{M}^B_{n-2}&=&\frac{1}{(n+1)^2} \langle\langle O_n\rangle\rangle \\
M_n^C&=&\frac{(n-1)n}{(n+1)^2}\langle\langle O_n\rangle\rangle \label{ltmom5}
\eq

There is another method to get the leading (theoretical-)twist part of the
distributions \ref{tpmasscor}, giving directly integrated equations
between the different wave functions. The different relations for
non-local operators in appendix \ref{app:oprel} are used : they give integrated forms for the leading
twist part of two-point operators.

One takes equation \ref{BBformula} :

\be
\left[\ov{\psi}(x)\gamma_\alpha\psi(-x)\right]_{\text{sym.}}=\int_0^1 du
\frac{\partial}{\partial x_\alpha}\ov{\psi}(ux)\hat{x}\psi(-ux) \label{vopsym}
\ee
where $\hat{x}=x^\nu\gamma_\nu$. The subscript "sym." means that in the Operator Product Expansion (local expansion around $x=0$), only the operators which have
symmetrical Lorentz-indices are taken. But for the leading theoretical twist, the traces has also to be removed, so the leading twist part of the right hand side of \ref{vopsym} is also taken :
\be
\left[\ov{\psi}(x)\gamma_\alpha\psi(-x)\right]_{\text{l. t.}}=\int_0^1 du
\frac{\partial}{\partial x_\alpha}\left[\ov{\psi}(ux)\hat{x}\psi(-ux) \right]_{\text{l. t.}}\label{voplt}
\ee

The condition for the leading twist part of $\ov{\psi}(ux)\hat{x}\psi(-ux)$ is the following (equation \ref{ltxhatapp}) :
\be
\frac{\partial}{\partial x_\alpha\partial x^\alpha}\left[\ov{\psi}(x)\hat{x}
\psi(-x)\right]_{\text{l. t.}}=0 \label{ltxhat}
\ee
 
Now the matrix element of $\ov{\psi}(ux)\hat{x}\psi(-ux)$ can be taken. The parameterization of the leading twist part up to $m_\rho^2$ corrections is the following :

\bq
\lefteqn{\langle 0|\left[\bar u(x) \hat{x} [x,-x] 
d(-x)\right]_{\text{l. t.}}|\rho^-(P,\lambda)\rangle =} \nonumber \\
&=&(\epl\cdot x)\int_0^1du g(u)\left[e^{i\xi Px}+\mx f_\xi(Px)+O(m_\rho^4)\right]
\label{xhatpar}
\eq
where $g(u)$ is the leading twist wave functions in the limit $m_\rho\rightarrow
 0$.

The condition \ref{ltxhat} can be applied; it gives (after some calculation) :
\be
\int_0^1 du g(u)\left[3f_\xi(Px)+(Px)f'_\xi(Px)-\xi^2e^{i\xi Px}\right]=0
\label{condfxi}
\ee

The following ansatz for $f_\xi(Px)$ can be done (having in mind the case of scalar field in the previous subsection \ref{sbscalar}) :
\be
f_\xi=a(\xi)\int_0^1b(v)e^{i\xi vPx}
\ee

Then equation \ref{condfxi} gives
\bq
a(\xi)&=&\xi^2 \\
b(v) &=&v^2 
\eq

This can be written in this form :
\bq
\lefteqn{\langle 0|\left[\bar u(x) \hat{x} [x,-x] 
d(-x)\right]_{\text{l. t.}}|\rho^-(P,\lambda)\rangle =} \nonumber \\
&=&\int_0^1du g(u)\left[(\epl\cdot x)e^{i\xi Px}\right]_{\text{l. t.}}
\eq
with
\bq
\lefteqn{\left[(\epl\cdot x)e^{i\xi Px}\right]_{\text{l. t.}}=} \nonumber \\
&=&(\epl\cdot x)\left[e^{i\xi Px}+\mx\xi^2\int_0^1 dv v^2 e^{iv\xi Px}\right]
\eq

The equation \ref{voplt} can be applied. After some algebra, one gets the distributions $A^{(e)}(u)$, $B^{(e)}(u)$, $C^{(e)}(u)$, $\hat{A}^{(e)}(u)$ and
$\hat{B}^{(e)}(u)$ defined in \ref{tpmasscor} in the leading (theoretical-) twist approximation :
\bq
A^{(e)}(u)&=&g(u) \label{Aelt}\\
\left.\int_0^1duB^{(e)}(u)e^{i\xi Px}\right)_{\text{l. t.}}&=&\int_0^1du g(u)\int_0^1dte^{it\xi Px} \\
\left.\int_0^1duC^{(e)}(u)e^{i\xi Px}\right)_{\text{l. t.}}&=&\int_0^1dug(u)\left[e^{i\xi Px}-3\int_0^1dt
e^{it\xi Px}+2\int_0^1dt\int_0^1dve^{ivt\xi Px}\right] \nonumber \\ \\
\left.\int_0^1du\hat{A}^{(e)}(u)e^{i\xi Px}\right)_{\text{l. t.}}&=&
\int_0^1dug(u)\xi^2\int_0^1dt\,t^2\left[e^{it\xi Px}-2\int_0^1dvv^2e^{ivt\xi Px}
\right] \\
\left.\int_0^1du\hat{B}^{(e)}(u)e^{i\xi Px}\right)_{\text{l. t.}}&=&
\int_0^1dug(u)\int_0^1dt\xi^2\int_0^1v^2dve^{ivt\xi Px} \label{Bhatelt}
\eq

There is no index "l. t." for $A^{(e)}(u)$ because this distribution contains only the leading twist part of the matrix element $\langle 0|\bar u(x) \gamma_\mu [x,-x] 
d(-x)|\rho^-(P,\lambda)\rangle$. So these last equations show that all the distributions can be computed when one has $A^{(e)}(u)$, in the leading (theoretical-)twist approximation. They reproduced exactly the relations between moments (equations \ref{ltmom1} to \ref{ltmom5}). The advantage is that they are directly in an integrated form.

%

\subsubsection{Adding twist 3 and twist 4 terms}

The idea is to add contributions of (theoretical-)twist 3 and 4 without the gluonic contribution. In the appendix \ref{app:oprel}, there is the following operator relation (\ref{dertot}) :
\bq
\ov \psi(x)\hat{x}\psi(-x)&=&\left[\ov \psi(x)\hat{x}\psi(-x)\right]_
{\text{l. t.}}-\frac{1}{4}x^2\int_0^1dv v^2\partial^2
\ov \psi(vx)\hat{x}\psi(-vx) \nonumber \\
& &{}+ \text{operators with gluons}+\ldots \label{txtdertot}
\eq

The matrix element of the term with a total derivative, gives
\bq
\lefteqn{
-\frac{1}{4}x^2\int_0^1dv v^2\langle 0|\partial^2\ov u(vx)\hat{x}[vx,-vx]d(-vx)|
\rho(P,\lambda)\rangle =} \nonumber \\
&=&m^2_\rho\frac{1}{4}x^2\int_0^1dv v^2\langle 0|\ov u(vx)\hat{x}[vx,-vx]d(-vx)|
\rho(P,\lambda)\rangle
\eq

The $O(m_\rho^0)$ term (equation \ref{xhatpar}) in the parameterization of the matrix element in the right hand side implies :
\bq
\lefteqn{
-\frac{1}{4}x^2\int_0^1dv v^2\langle 0|\partial^2\ov u(vx)\hat{x}[vx,-vx]d(-vx)|
\rho(P,\lambda)\rangle =} \nonumber \\
&=&(\epl\cdot x)\mx\int_0^1 du g(u)\int_0^1dv v^2e^{iv\xi Px}
\eq

So the leading twist part with the total derivative term is :
\bq
\lefteqn{\langle 0|\left[\bar u(x) \hat{x} [x,-x] 
d(-x)\right]_{\text{l. t. + tot. der.}}|\rho^-(P,\lambda)\rangle =} \nonumber \\
&=&\int_0^1du g(u)\left[(\epl\cdot x)e^{i\xi Px}\right]_
{\text{l. t. + tot. der.}}
\eq
with
\bq
\lefteqn{\left[(\epl\cdot x)e^{i\xi Px}\right]_{\text{l. t. + tot. der.}}=} \nonumber \\
&=&(\epl\cdot x)\left[e^{i\xi Px}+\mx(\xi^2+1)\int_0^1 dv v^2 e^{iv\xi Px}\right]
\label{totdercor}
\eq

This last form can be used to give some twist 4 corrections to equation \ref{Aelt}
to \ref{Bhatelt}.

Now the twist 3 part without gluon can be added. The matrix element of the equations \ref{twist3} is taken, suppressing the gluonic terms :
\bq
\lefteqn{\langle 0|\left[\bar u(x) \gamma_\mu [x,-x] 
d(-x)\right]_{\text{t3}}|\rho^-(P,\lambda)\rangle =} \nonumber \\ 
&=&-i\epsilon_{\mu\nu\alpha\beta}\int_0^1dt\,t x^\nu(-iP^\alpha)\langle 0|\bar u(tx) \gamma_\beta\gamma_5 [tx,-tx] 
d(-tx)|\rho^-(P,\lambda)\rangle \nonumber \\ \\
\lefteqn{\langle 0|\left[\bar u(x) \gamma_\mu \gamma_5[x,-x] 
d(-x)\right]_{\text{t3}}|\rho^-(P,\lambda)\rangle =} \nonumber \\ 
&=&-i\epsilon_{\mu\nu\alpha\beta}\int_0^1dt\,t x^\nu(-iP^\alpha)\langle 0|\bar u(tx) \gamma_\beta [tx,-tx] 
d(-tx)|\rho^-(P,\lambda)\rangle \nonumber \\
\eq

Combining these two equations, replacing the right hand side of the second one with the
parameterization of \ref{tpmasscor}, one gets :
\bq
\lefteqn{\langle 0|\left[\bar u(x) \gamma_\mu [x,-x] 
d(-x)\right]_{\text{t3}}|\rho^-(P,\lambda)\rangle =} \nonumber \\ 
&=&(\epl_\mu(P\cdot x)^2+P_\mu(\epl\cdot x)(P\cdot x)\int_0^1du \int_0^1 dt\,t
\int_0^1 dv v^2 B^{(e)}(u)e^{ivt\xi Px} \nonumber \\
& &{}+m^2_\rho\int_0^1du\int_0^1 dt t\int_0^1 dv v^2 e^{ivt\xi Px}\times
\nonumber \\
& &{}\;\times \left[(\epl_\mu(P\cdot x)^2+P_\mu(\epl\cdot x)(P\cdot x)
\frac{t^2x^2v^2}{4}\hat{B}^{(e)}(u)+\left(x^2\epl_\mu-x_\mu(\epl\cdot x)\right)
B^{(e)}(u)\right] \nonumber \\
\eq

Including the leading twist term, the total derivative term (\ref{totdercor}) and this last equation, equations for $A^{(e)}(u)$, $B^{(e)}(u)$, $C^{(e)}(u)$, $\hat{A}^{(e)}(u)$ and
$\hat{B}^{(e)}(u)$ can be obtained :
\bq
A^{(e)}(u)&=&g(u) \label{Aet234}\\
\int_0^1duB^{(e)}(u)e^{i\xi Px}&=&\int_0^1du g(u)\int_0^1dte^{it\xi Px}\nonumber
\\
& &{}-(P\cdot x)^2\int_0^1du\int_0^1dt\,t\int_0^1dv\,v^2e^{ivt\xi Px}B^{(e)}(u)
\nonumber \\ \\
-\frac{1}{2}\frac{1}{(P\cdot x)^2}\int_0^1duC^{(e)}(u)e^{i\xi Px}&=&\int_0^1du\frac{\xi^2+1}{\xi}g(u)\int_0^1dt\,t^2\int_0^1dvv^2e^{ivt\xi Px} \nonumber \\ 
& &{}-\int_0^1du B^{(e)}(u)\int_0^1dt\,t\int_0^1dv\,v^2e^{ivt\xi Px} \nonumber \\
\\
\int_0^1du\hat{B}^{(e)}(u)e^{i\xi Px}&=&
\int_0^1dug(u)(\xi^2+1)\int_0^1dt\,t^2\int_0^1v^2dve^{ivt\xi Px} \nonumber \\
& &{}+4\int_0^1duB^{(e)}(u)\int_0^1dt\,t\int_0^1dv\,v^2e^{ivt\xi Px} \nonumber
\\
& &{}-(P\cdot x)^2\int_0^1du\hat{B}^{(e)}(u)\int_0^1dt\,t^3
\int_0^1dv\,v^3e^{itv\xi Px} \nonumber \\ \label{Bhatet234}
\eq
\bq\lefteqn{\int_0^1du\hat{A}^{(e)}(u)e^{i\xi Px}=} \nonumber \\
& &{}
\int_0^1dug(u)(\xi^2+1)\int_0^1dt\,t^2\int_0^1dv\,v^2\left(i\xi(P\cdot x)+1\right)
t^2e^{ivt\xi Px} \nonumber \\
& &{}+4\int_0^1duB^{(e)}(u)\int_0^1dt\,t\int_0^1dv\,v^2e^{ivt\xi Px} 
\nonumber \\
\eq

Theoretically, all the chiral-even distributions $B^{(e)}(u)$, $C^{(e)}(u)$, $\hat{A}^{(e)}(u)$ can be computed once the leading (physical-)twist contribution $A^{(e)}(u)$ is known. But, practically, it quite long and hard if one has a conformal expansion for $A^{(e)}(u)$ (except if only the asymptotic part of the wave function is taken, see \cite{BBS}). The advantage of these equations is that they are independent of the model used to computed $A^{(e)}(u)$ (another expansion can be used as the conformal one). They can be the basis of a numerical computation if one has discrete values for $A^{(e)}(u)$.

In the appendix \ref{app:masscorr}, similar equations for the chiral-odd distributions are given.

%

\subsection{Mass correction using the equation of motion}

Equation of motion is used to relate the two-point wave functions to the three points ones (these calculations come from \cite{BBS}). 

From \cite{BF}, page 245, one has :
\be
i\partial_\mu\ov u(x)\gamma^\mu d(-x)=\int_{-1}^1dv\ov u(x)gG_{\mu\nu}
(vx)\gamma^\nu d(-x)
\ee
The symbol $\partial_\mu$ is the derivative over the total translation define in \ref{totder}.

When the matrix element $\langle 0|\ldots|\rho(P,\lambda)\rangle$ is taken (replacing $x$ by $z$), on gets (using the definitions \ref{eq:vda} and \ref{chie3} for $\phi_\|(u)$, $g_3(u)$, $\Phi(u)$ and $\Psi(u)$)  :
\bq
\lefteqn{\frac{1}{2}m_\rho^3 \frac{\epl\cdot z}{p\cdot z}\int_0^1due^{i\xi pz}
\left[\phi_\| (u)-g_3(u)\right]=} \nonumber \\
&=&-if_\rho m_\rho^3(\epl\cdot z)\int_{-1}^1dv\int {\cal D}\un{\alpha}
\left[2\Phi(\un{\alpha})+\Psi(\un{\alpha})\right]
e^{-ipz(\alpha_u-\alpha_d+v\alpha_g)}
\nonumber \\
\eq
 
The integration of the right hand side can be transformed :
\be
\int_{-1}^1 dv\int{\cal D}\un \alpha F(\un \alpha)e^{-ipz(\alpha_u-\alpha_d+
v\alpha_g)}=2\int_0^1dwe^{ipz(2w-1)}\int_0^wd\alpha_d\int_0^{1-x}d\alpha_u
\frac{1}{\alpha_g}F(\un \alpha)
\ee

And a relation between the two-points and three-points distributions is obtained :
\be
\frac{1}{2}\left[\phi_\|(u)-g_3(u)\right]=\frac{d}{du}\int_0^ud\alpha_d
\int_0^{1-u}d\alpha_u\frac{1}{\alpha_g}\left[2\Phi(\un \alpha)
+\Psi(\un \alpha)\right]
\ee

In a similar way, the operator relation (in \cite{BF})
\be
\frac{\partial}{\partial x_\mu}\ov u(x)\gamma_\mu d(-x) = -i\int_{-1}^1dv\,v
\ov u(x)x^\alpha g G_{\alpha\beta}(vx)\gamma^\beta d(-x)
\ee
gives
\be
\frac{1}{8}\hat{A}^{(e)}(u)+2C_2^{(e)}(u)=\int_0^ud\alpha_d\int_0^{1-u}d\alpha_u
\left(\alpha_uu-\alpha_d(1-u)\right)\frac{1}{\alpha_g^2}\left[2\Phi(\un{\alpha})
+\Psi(\un{\alpha})\right]
\ee
with
\be
C_2^{(e)}(u)=\int_0^u\int_0^v\;dtC^{(e)}(t)
\ee

Hence from the wave functions computed in section \ref{PsiPhi}, the mass correction of the two-points functions can be computed. The advantage of this method is that it is quite simple, but it depends on the model for $\Phi$ and $\Psi$, and there is no general methods to find equations that relate the two-point to the three points functions.


\sect{Conclusion}

In this work, I describe the different wave functions of meson, the method conformal expansion and its application to the non-leading twist wave functions. I also give methods of computing the mass correction, which helps to apply these results to heavier vector mesons as the $\rho$.

The use of conformal expansion and exact operators relations coming from the equation of motion is still long and heavy, a more general method would be certainly useful, and there is no work of higher gluonic corrections, like four-points wave functions for example.  

These non-local, non-perturbative objects are needed for the computation of exclusive decays, but they also describe the nature of the quark confinement in meson in a Lorentz-invariant and gauge invariant way. A better knowledge of the non-leading twist wave functions would probably help to understand how QCD describes the physics of hadron.

Another question is the possible link to effective low energy theories; for example, the quark condensate $\langle 0|\ov{q}q|0\rangle$ is an order parameter of the breakdown of the chiral symmetry at low energy; can one found such thing with non-local wave functions ?

The basis of the conformal expansion is the conformal invariance of QCD to one loop. But a analysis of the two-loop effects on this expansion, a better renormalization group dependence of these wave functions would be interesting.

Other ways of determining these wave functions (by experiments or using another expansion as the conformal one) is certainly desirable, and the "theoretical twist method" of determining the mass corrections that I give in this work will still be applicable. 

\section*{Acknowledgments}

This work is supported by Schweizerischer Nationalfond. I am grateful to V. M. Braun and P. Ball for introducing me to the subject and to D. Wyler for encouragement and for proofreading this work.


\begin{appendix}
\sect{List of vector-meson wave functions up to twist $4$}

In this appendix, I give a list of vector-meson wave functions (the chosen meson is the $\rho$), following the notation of \cite{BBKT} and \cite{BBS}.

%
%

\subsection{Two-points distributions}
Chiral-odd $\rho$ two-point distributions :
\begin{eqnarray}
\lefteqn{\hspace*{-1.5cm}\langle 0|\bar u(z) \sigma_{\mu \nu} [z,-z] 
d(-z)|\rho^-(P,\lambda)\rangle =} \nonumber \\
&=& i f_{\rho}^{T} \left[ ( e^{(\lambda)}_{\perp \mu}p_\nu -
e^{(\lambda)}_{\perp \nu}p_\mu )
\int_{0}^{1} \!du\, e^{i \xi p \cdot z} \phi_{\perp}(u, \mu^{2}) \right. 
\nonumber \\
& &{}+ (p_\mu z_\nu - p_\nu z_\mu )
\frac{e^{(\lambda)} \cdot z}{(p \cdot z)^{2}}
m_{\rho}^{2} 
\int_{0}^{1} \!du\, e^{i \xi p \cdot z} \htt (u, \mu^{2}) 
\nonumber \\
& & \left.{}+ \frac{1}{2}
(e^{(\lambda)}_{\perp \mu} z_\nu -e^{(\lambda)}_{\perp \nu} z_\mu) 
\frac{m_{\rho}^{2}}{p \cdot z} 
\int_{0}^{1} \!du\, e^{i \xi p \cdot z} h_{3}(u, \mu^{2}) \right],
\label{eq:tda}
\end{eqnarray}
\begin{equation}
\langle 0|\bar u(z) [z,-z] 
d(-z)|\rho^-(P,\lambda)\rangle
= -i \left(f_{\rho}^{T} - f_{\rho}\frac{m_{u} + m_{d}}{m_{\rho}}
\right)(e^{(\lambda)}\cdot z) m_{\rho}^{2}
\int_{0}^{1} \!du\, e^{i \xi p \cdot z} \hs(u, \mu^{2}),
\label{eq:sda}
\end{equation}

\noindent Chiral-even two-points distributions :
\begin{eqnarray}
\lefteqn{\langle 0|\bar u(z) \gamma_{\mu} [z,-z] 
d(-z)|\rho^-(P,\lambda)\rangle = }\makebox[2cm]{\ } \nonumber \\
&=& f_{\rho} m_{\rho} \left[ p_{\mu}
\frac{e^{(\lambda)}\cdot z}{p \cdot z}
\int_{0}^{1} \!du\, e^{i \xi p \cdot z} \phi_{\parallel}(u, \mu^{2}) \right. 
+ e^{(\lambda)}_{\perp \mu}
\int_{0}^{1} \!du\, e^{i \xi p \cdot z} g_{\perp}^{(v)}(u, \mu^{2}) 
\nonumber \\
& & \left.{}- \frac{1}{2}z_{\mu}
\frac{e^{(\lambda)}\cdot z }{(p \cdot z)^{2}} m_{\rho}^{2}
\int_{0}^{1} \!du\, e^{i \xi p \cdot z} g_{3}(u, \mu^{2}) \right]
\label{eq:vda}
\end{eqnarray}
and 
\begin{eqnarray}
\lefteqn{\langle 0|\bar u(z) \gamma_{\mu} \gamma_{5}[z,-z] 
d(-z)|\rho^-(P,\lambda)\rangle = }\makebox[2cm]{\ } \nonumber \\
&=& \frac{1}{2}\left(f_{\rho} - f_{\rho}^{T}
\frac{m_{u} + m_{d}}{m_{\rho}}\right)
m_{\rho} \epsilon_{\mu}^{\phantom{\mu}\nu \alpha \beta}
e^{(\lambda)}_{\perp \nu} p_{\alpha} z_{\beta}
\int_{0}^{1} \!du\, e^{i \xi p \cdot z} g^{(a)}_{\perp}(u, \mu^{2}).
\label{eq:avda}
\end{eqnarray}
with the notation
$$\xi = u - (1-u) = 2u-1.$$ 

The vector and tensor  decay constants $f_\rho$ and $f_\rho^T$ are defined 
as usually as
\begin{eqnarray}
\langle 0|\bar u(0) \gamma_{\mu}
d(0)|\rho^-(P,\lambda)\rangle & = & f_{\rho}m_{\rho}
e^{(\lambda)}_{\mu},
\label{eq:fr}\\
\langle 0|\bar u(0) \sigma_{\mu \nu} 
d(0)|\rho^-(P,\lambda)\rangle &=& i f_{\rho}^{T}
(e_{\mu}^{(\lambda)}P_{\nu} - e_{\nu}^{(\lambda)}P_{\mu}),
\label{eq:frp}
\end{eqnarray}

\begin{table}
\begin{center}
\renewcommand{\arraystretch}{1.3}
\begin{tabular}{|c|ccc|}
\hline
Twist    & 2 & 3 & 4 \\
    & $O(1)$  & $O(1/Q)$& $O(1/Q^{2})$ \\ \hline
$ $ & $\phi_{\parallel}$ & $\htt$, $\hs$& 
$g_{3}$ \\
$ $ & $\phi_{\perp}$ & $g_{\perp}^{(v)}$,
$g_{\perp}^{(a)}$ & $h_{3}$\\[2pt] \hline
\end{tabular}
\renewcommand{\arraystretch}{1}
\end{center}
\caption{"Physical twist" classification of two-points distributions}
\label{tab:2}
\end{table}

If every $m^2_\rho$ terms are needed, $x^2$ corrections has to be made. In that case, the notation is a little different :

\noindent Chiral-odd two-points distributions :
\begin{eqnarray}
\lefteqn{\hspace*{-1.5cm}\langle 0|\bar u(x) \sigma_{\mu \nu} [x,-x] 
d(-x)|\rho^-(P,\lambda)\rangle =} \nonumber \\
&=& i f_{\rho}^{T} \left[ ( e^{(\lambda)}_{\mu}P_\nu -
e^{(\lambda)}_{\nu}P_\mu )
\int_{0}^{1} \!du\, e^{i \xi P \cdot x} \left(A^{(o)}(u) 
+\frac{m^2_\rho x^2}{4}\hat{A}^{(o)}(u)\right)\right. 
\nonumber \\
& &{}+ (P_\mu x_\nu - P_\nu x_\mu )
\frac{e^{(\lambda)} \cdot x}{(P \cdot x)^{2}}m_{\rho}^{2} 
\int_{0}^{1} \!du\, e^{i \xi P \cdot x}\left(B^{(o)}(u) 
+\frac{m^2_\rho x^2}{4}\hat{B}^{(o)}(u)\right)  
\nonumber \\
& & \left.{}+ \frac{1}{2}
(e^{(\lambda)}_{\mu} x_\nu -e^{(\lambda)}_{\nu} x_\mu) 
\frac{m_{\rho}^{2}}{P \cdot x} 
\int_{0}^{1} \!du\, e^{i \xi P \cdot x} C^{(o)}(u) \right],
\label{eq:frpm}
\end{eqnarray}
with
\bq
A^{(o)}(u)&=&\phi_\perp(u) \\
B^{(o)}(u)&=&h_\|^{(t)}(u)-\frac{1}{2}\phi_\perp(u)-\frac{1}{2}h_3(u) \\
C^{(o)}(u)&=&h_3(u)-\phi_\perp(u)
\eq

\noindent Chiral-even two-points distributions :
\begin{eqnarray}
\lefteqn{\langle 0|\bar u(x) \gamma_{\mu} [x,-x] 
d(-x)|\rho^-(P,\lambda)\rangle = }\makebox[2cm]{\ } \nonumber \\
&=& f_{\rho} m_{\rho} \left[ P_{\mu}
\frac{e^{(\lambda)}\cdot x}{P \cdot x}
\int_{0}^{1} \!du\, e^{i \xi P \cdot x} \left(A^{(e)}(u)+
\frac{x^2 m_\rho^2}{4}\hat{A}^{(e)}(u)\right) \right. 
\nonumber \\
& &{}+ \left(e^{(\lambda)}_{\mu}-P_\mu\frac{\epl\cdot x}{P\cdot x}\right)
\int_{0}^{1} \!du\, e^{i \xi P \cdot x} \left(B^{(e)}(u) 
+\frac{x^2 m_\rho^2}{4}\hat{B}^{(e)}(u)\right)\nonumber \\
& & \left.{}- \frac{1}{2}x_{\mu}
\frac{e^{(\lambda)}\cdot x }{(P \cdot x)^{2}} m_{\rho}^{2}
\int_{0}^{1} \!du\, e^{i \xi P \cdot x} C^{(e)}(u) \right]
\label{eq:vdam}
\end{eqnarray}
with
\bq
A^{(e)}(u)&=&\phi^\|(u) \\
B^{(e)}(u)&=&g^{(v)}_\perp(u) \\
C^{(u)}(u)&=&g_3(u)+\phi_\|(u)-2g^{(v)}_\perp(u)
\eq

%
%

\subsection{Three-points distributions}

\noindent Chiral-odd three-points distributions :

\bq
\lefteqn{\langle 0|\bar u(z) \sigma_{\alpha\beta} [z,vz]
         gG_{\mu\nu}(vz)[vz,-z] 
         d(-z)|\rho^-(P,\lambda)\rangle =}\makebox[2cm]{\ } \nonumber \\
&=& f_\rho^T m_\rho^2 \left[\frac{e^{(\lambda)}\cdot z }{2 (p \cdot z)}
    [ p_\alpha p_\mu g^\perp_{\beta\nu} 
     -p_\beta p_\mu g^\perp_{\alpha\nu} 
     -p_\alpha p_\nu g^\perp_{\beta\mu} 
     +p_\beta p_\nu g^\perp_{\alpha\mu} ]   
      f_{3\rho}^T m_{\rho} {\cal T}(v,pz) \right. \nonumber \\
& &{}+
	[ p_\alpha\epl_{\perp\mu} g^\perp_{\beta\nu}
	 -p_\beta\epl_{\perp\mu} g^\perp_{\alpha\nu}
	 -p_\alpha\epl_{\perp\nu} g^\perp_{\beta\mu}
	 +p_\beta\epl_{\perp\nu} g^\perp_{\alpha\mu} ]
	{\cal T}_1(v,pz)\nonumber \\
& &{}+
	[ p_\mu\epl_{\perp\alpha} g^\perp_{\beta\nu}
	 -p_\mu\epl_{\perp\beta} g^\perp_{\alpha\nu}
	 -p_\nu\epl_{\perp\alpha} g^\perp_{\beta\mu}
	 +p_\nu\epl_{\perp\beta} g^\perp_{\alpha\mu} ]
	{\cal T}_2(v,pz) \nonumber \\
& &{}+  
	[ p_\alpha\epl_{\perp\beta} p_\mu z_\nu
	 -p_\beta\epl_{\perp\alpha} p_\mu z_\nu
	 -p_\alpha\epl_{\perp\beta} p_\nu z_\mu
	 +p_\beta\epl_{\perp\alpha} p_\mu z_\nu ]
	{\cal T}_3(v,pz) \nonumber \\
& &{}+\left.
	[ p_\alpha\epl_{\perp\nu} p_\mu z_\beta
	 -p_\beta\epl_{\perp\nu} p_\mu z_\alpha
	 -p_\alpha\epl_{\perp\mu} p_\nu z_\beta
	 +p_\beta\epl_{\perp\mu} p_\nu z_\alpha ]
	{\cal T}_4(v,pz) \right] \nonumber \\
\eq
where ${\cal T}$ is a twist 3 distribution ("physical twist"), the other are twist 4. There are other
Lorentz-structures which are twist 5.

\bq
\lefteqn{\langle 0|\bar u(z)  [z,vz]
         gG_{\mu\nu}(vz)[vz,-z] 
         d(-z)|\rho^-(P,\lambda)\rangle =}\makebox[2cm]{\ } \nonumber \\
&=& i f^T_\rho m_\rho^2[\epl_{\perp\mu}p_\nu -e_{\perp\nu}]S(v,pz)
\eq

\bq
\lefteqn{\langle 0|\bar u(z)  [z,vz] i \gamma_5
         g\Gtilde_{\mu\nu}(vz)[vz,-z] 
         d(-z)|\rho^-(P,\lambda)\rangle =}\makebox[2cm]{\ } \nonumber \\
&=& i f^T_\rho m_\rho^2[\epl_{\perp\mu}p_\nu -e_{\perp\nu}]\tilde{S}(v,pz)
\eq

These $S(v,pz)$ and $\tilde{S}(v,pz)$ are twist 4 distributions.

\noindent Chiral-even three-points distributions :
\bq
\lefteqn{\langle 0|\bar u(z) i\gamma_\alpha [z,vz]gG_{\mu\nu}(vz)[vz,-z] 
	d(-z)|\rho^-(P,\lambda)\rangle=}\makebox[2cm]{\ }\nonumber \\
&=&   f_{\rho}m_\rho\left[ p_\alpha[p_\mu e^{(\lambda)}_{\perp\nu}-p_\nu
	e^{(\lambda)}_{\perp\mu}]{\cal V}(v,pz)\right. \nonumber \\
& &{}+m_\rho^2\frac{\epl\cdot z}{p\cdot z}[p_\mu g_{\alpha\nu}^\perp
	-p_\nu g_{\alpha\mu}^\perp]\Phi(v,pz) \nonumber \\
& &{}\left.+m_\rho^2\frac{\epl\cdot z}{(p\cdot z)^2}[p_\mu z_\nu-p_\nu z_\mu]
	p_\alpha \Psi(v,pz) \right] \label{chie3}
\eq
\bq
\lefteqn{\langle 0|\bar u(z) \gamma_\alpha \gamma_5
        [z,vz]g\Gtilde_{\mu\nu}(vz)[vz,-z] 
	d(-z)|\rho^-(P,\lambda)\rangle=}\makebox[2cm]{\ }\nonumber \\
&=&if_{\rho}m_\rho\left[ p_\alpha[p_\mu e^{(\lambda)}_{\perp\nu}-p_\nu        	
	e^{(\lambda)}_{\perp\mu}]{\cal A}(v,pz)\right. \nonumber \\
& &{}+m_\rho^2\frac{\epl\cdot z}{p\cdot z}[p_\mu g_{\alpha\nu}^\perp
	-p_\nu g_{\alpha\mu}^\perp]\tilde{\Phi}(v,pz) \nonumber \\
& &{}\left.+m_\rho^2\frac{\epl\cdot z}{(p\cdot z)^2}[p_\mu z_\nu-p_\nu z_\mu]
	p_\alpha \tilde{\Psi}(v,pz) \right]
\eq
where ${\cal A}(v,pz)$ and ${\cal V}(v,pz)$ are twist 3 distributions, the other twist 4.

Every functions ${\cal F}(v,pz)$ have such form : 
\begin{equation}
   {\cal F}(v,pz) =\int {\cal D}\underline{\alpha} 
e^{-ipz(\alpha_u-\alpha_d+v\alpha_g)}{\cal F}(\underline{\alpha}),
\end{equation}
and $\underline{\alpha}$ is the set of three momentum fractions
$\underline{\alpha}=\{\alpha_d,\alpha_u,\alpha_g\}$.
 The integration measure is defined as 
\begin{equation}
 \int {\cal D}\underline{\alpha} \equiv \int_0^1 d\alpha_d
  \int_0^1 d\alpha_u\int_0^1 d\alpha_g \,\delta(1-\sum \alpha_i).
\label{eq:measure}
\end{equation}

%
%
%
%
\sect{Models for $\rho$ wave functions up to twist 4}

I reproduce
the results of \cite{BB} here, which helps to understand section \ref{PsiPhi}. Their computations can be found in \cite{BBKT} and \cite{BBS}, where a more complete list can be found.

%
%

\subsection{Chiral-even distributions}

The model for the leading twist 2 distribution amplitude $\phi_\parallel$ is
\begin{equation}\label{eq:phipar}
\phi_\parallel(u) =  6 u\bar u \left[ 1 + 3 a_1^\parallel\, \xi +
a_2^\parallel\, \frac{3}{2} ( 5\xi^2  - 1 ) \right]
\end{equation}
with parameter values as specified in Tab.~\ref{tab:para}.
The expressions for higher-twist distributions given below correspond
to the simplest self-consistent approximation which satisfies the  
QCD equations of motion \cite{BBKT,BBS} :
\begin{itemize}
\item{} Three-particle distributions of twist~3 :
\begin{eqnarray}
{\cal V} (\underline{\alpha}) &=& 
540\, \zeta_3 \omega^V_3 (\alpha_d-\alpha_u)\alpha_d \alpha_u\alpha_g^2,
\label{modelV}\\
{\cal A} (\underline{\alpha}) &=& 
360\,\zeta_3 \alpha_d \alpha_u \alpha_g^2 
\Big[ 1+ \omega^A_{3}\frac{1}{2}(7\alpha_g-3)].
\label{modelA}
 \end{eqnarray}
\item{} Two-particle distributions of twist~3 :
\begin{eqnarray}
g_\perp^{(a)}(u) & = & 6 u \bar u \left[ 1 + a_1^\parallel \xi +
\left\{\frac{1}{4}a_2^\parallel +
\frac{5}{3}\, \zeta_{3} \left(1-\frac{3}{16}\,
\omega^A_{3}+\frac{9}{16}\omega^V_3\right)\right\}
(5\xi^2-1)\right]\nonumber\\
& & {} + 6\, \widetilde{\delta}_+ \,  (3u \bar u + \bar u \ln \bar u +
u \ln u ) + 
6\, \widetilde{\delta}_- \,  (\bar u \ln \bar u - u \ln u),\\
 g_\perp^{(v)}(u) & = & \frac{3}{4}(1+\xi^2)
+ a_1^\parallel\,\frac{3}{2}\, \xi^3 
 + \left(\frac{3}{7} \, 
a_2^\parallel + 5 \zeta_{3} \right) \left(3\xi^2-1\right)
 \nonumber\\
& & {}+ \left[ \frac{9}{112}\, a_2^\parallel 
+ \frac{15}{64}\, \zeta_{3}\Big(3\,\omega_{3}^V-\omega_{3}^A\Big)
 \right] \left( 3 - 30 \xi^2 + 35\xi^4\right)\nonumber\\
& & {}+\frac{3}{2}\,\widetilde{\delta}_+\,(2+\ln u + \ln\bar u) +
\frac{3}{2}\,\widetilde{\delta}_-\, ( 2 \xi + \ln\bar u - \ln u),\label{eq:gv}
\end{eqnarray}
\item{} Three-particle distributions of twist~4 :
\begin{eqnarray}
\widetilde \Phi (\underline{\alpha}) &=& 
  \Big[-\frac{1}{3}\zeta_{3}+\frac{1}{3}\zeta_{4}\Big] 
   30(1-\alpha_g)\alpha_g^2,
\nonumber\\
  \Phi (\underline{\alpha}) &=& 
  \Big[-\frac{1}{3}\zeta_{3}+\frac{1}{3}\zeta_{4}\Big] 
   30(\alpha_u-\alpha_d)\alpha_g^2,
\nonumber\\
  \widetilde\Psi (\underline{\alpha}) &=& 
  \Big[\frac{2}{3}\zeta_{3}+\frac{1}{3}\zeta_{4}\Big] 
   120 \alpha_u\alpha_d\alpha_g,
\nonumber\\
 \Psi (\underline{\alpha}) &=& 0.
\end{eqnarray}
\item{} Two-particle distributions of twist~4 : 
\begin{eqnarray}
A^{(e)}(u) &=& \Bigg[\frac{4}{5}+\frac{20}{9} \zeta_{4}
                    +\frac{8}{9} \zeta_{3}\Bigg]30 u^2(1-u)^2,
\nonumber\\
g_3(u) &=& 6u(1-u) + \Bigg[\frac{10}{3} \zeta_{4}
                    -\frac{20}{3} \zeta_{3}\Bigg](1-3 \xi^2),
\nonumber\\
C^{(e)}(u) &=& \Bigg[\frac{3}{2}+\frac{10}{3} \zeta_{4}
                    +\frac{10}{3} \zeta_{3}\Bigg](1-3 \xi^2),
\end{eqnarray}
\end{itemize} 
where the dimensionless couplings $\zeta_3$ and $\zeta_4$ are 
defined as local matrix elements
\begin{eqnarray}
\langle0|\bar u g\tilde G_{\mu\nu}\gamma_\alpha
 \gamma_5 d|\rho^-(P,\lambda)\rangle  &=&
f_\rho m_\rho \zeta_{3}
\Bigg[
e^{(\lambda)}_\mu\Big(P_\alpha P_\nu-\frac{1}{3}m^2_\rho \,g_{\alpha\nu}\Big)  
-e^{(\lambda)}_\nu\Big(P_\alpha P_\mu-\frac{1}{3}m^2_\rho \,g_{\alpha\mu}\Big)
\Bigg]
\nonumber\\
&&{}+\frac{1}{3}f_\rho m_\rho^3 \zeta_{4}
\Bigg[e^{(\lambda)}_\mu g_{\alpha\nu}- e^{(\lambda)}_\nu g_{\alpha\mu}\Bigg]  
\label{defxi}
\end{eqnarray}
and have been estimated from QCD sum rules 
 \cite{ZZC85,BK86}.
The terms in $\delta_\pm$ and $\widetilde\delta_\pm$ specify quark-mass 
corrections in twist~3 distributions induced by the equations of motion.
The numerical values of these and other coefficients are listed in 
Tabs.~\ref{tab:para} and \ref{tab:para2}\footnote{In the 
notations of Ref.~\cite{BBKT},
$\omega_{1,0}^A\equiv \omega_3^A$, $ \zeta_3^A\equiv \zeta_3$ and 
$\zeta_3^V \equiv (3/28)\zeta_3\omega_3^V$.}.
Note that SU(3) breaking effects are neglected in twist 4 distributions
and in gluonic parts of twist~3 distributions.

The numbers $a_1^\|$, $a_2^\|$, $\omega_3^A$, $\omega_3^V$ are computed in the appendix C of \cite{BBKT}, using usual two-points QCD sum rule. $\tilde\delta_+$ and $\tilde\delta_-$ have following definitions :
\be
\tilde\delta_\pm=\frac{f_\rho^T}{f_\rho}\frac{m_u\pm m_d}{m_\rho}
\ee

%
%

\subsection{Chiral-odd distributions}

\begin{table}
\renewcommand{\arraystretch}{1.4}
\addtolength{\arraycolsep}{3pt}
$$
\begin{array}{|c|cccc|}
\hline
V & \rho^\pm & K^*_{u,d} & \bar{K}^*_{u,d}& \phi\\ \hline
f_V [{\rm MeV}] & 198\pm 7  & 226 \pm 28& 226 \pm 28 & 254 \pm 3\\
f^T_V [{\rm MeV}] & 
\begin{array}{c} 160\pm 10 \\ 152\pm 9 \end{array}&
\begin{array}{c} 185\pm 10 \\ 175\pm 9 \end{array}&
\begin{array}{c} 185\pm 10 \\ 175\pm 9 \end{array}&
\begin{array}{c} 215\pm 15 \\ 204\pm 14 \end{array}
\\ \hline
a_1^\parallel & 0 & 
\begin{array}{c} 0.19 \pm 0.05 \\ 0.17\pm 0.04 \end{array}&
\begin{array}{c} -0.19 \pm 0.05 \\ -0.17\pm 0.04 \end{array}&
\phantom{-}0\\
a_2^\parallel & 
\begin{array}{c} 0.18 \pm 0.10 \\ 0.16\pm 0.09 \end{array}&
\begin{array}{c} 0.06 \pm 0.06 \\ 0.05\pm 0.05 \end{array}&
\begin{array}{c} \phantom{-}0.06 \pm 0.06 \\ 
                   \phantom{-}0.05\pm 0.05 \end{array}&
0\pm0.1\\
a_1^\perp & 0 & 
\begin{array}{c} 0.20 \pm 0.05 \\ 0.18\pm 0.05 \end{array}&
\begin{array}{c} -0.20 \pm 0.05 \\ -0.18\pm 0.05 \end{array}&
\phantom{-}0\\
a_2^\perp & 
\begin{array}{c} 0.20 \pm 0.10 \\ 0.17\pm 0.09 \end{array}&
\begin{array}{c} 0.04 \pm 0.04 \\ 0.03\pm 0.03 \end{array}&
\begin{array}{c} \phantom{-}0.04 \pm 0.04 \\ 
                  \phantom{-}0.03\pm 0.03 \end{array}&
 0\pm0.1\\ \hline
\delta_+ & 0 & 
\begin{array}{c} \phantom{-}0.24 \\ \phantom{-}0.22 \end{array}&
\begin{array}{c} 0.24 \\ 0.22 \end{array}&
\begin{array}{c} 0.46 \\ 0.41 \end{array}
\\
\delta_- & 0 & 
\begin{array}{c} -0.24 \\ -0.22 \end{array}&
\begin{array}{c} 0.24 \\ 0.22 \end{array}&
0 \\
\widetilde{\delta}_+ & 0 & 
\begin{array}{c} \phantom{-}0.16 \\ \phantom{-}0.13 \end{array}&
\begin{array}{c} 0.16 \\ 0.13 \end{array}&
\begin{array}{c} 0.33 \\ 0.27 \end{array}
\\
\widetilde{\delta}_- & 0 & 
\begin{array}{c} -0.16 \\ -0.13 \end{array}&
\begin{array}{c} 0.16 \\ 0.13 \end{array}&
0\\ \hline
\end{array}
$$
\caption[]{Masses and couplings of vector meson distribution
  amplitudes including SU(3) breaking. In cases that two values 
are given, the upper one corresponds to the scale $\mu^2=1\,$GeV$^2$ 
and the lower one to $\mu^2 = m_B^2-m_b^2 = 4.8\,$GeV$^2$,
respectively.
$m_s$ is taken at $\mu=1\,{\rm GeV}$ ($ms= 150\,{\rm MeV}$ and the quark masses $m_u$ and $m_d$ are put to zero).
}\label{tab:para}
$$
\begin{array}{|c|ccccccc|}\hline
& \zeta_3 & \omega_3^A & \omega_3^V & \omega_3^T & \zeta_4 & \zeta_4^T
& \tilde{\zeta_4^T}\\ \hline
V &
\begin{array}{c} 0.032\\ 0.023 \end{array}& 
\begin{array}{c} -2.1 \\  -1.8 \end{array}& 
\begin{array}{c} 3.8  \\   3.7 \end{array}& 
\begin{array}{c} 7.0  \\   7.5 \end{array}&
\begin{array}{c} 0.15 \\   0.13 \end{array}&  
\begin{array}{c} 0.10 \\   0.07 \end{array}&  
\begin{array}{c} -0.10\\  -0.07 \end{array}  
\\ \hline
\end{array}
$$
\caption[]{Couplings for twist 3 and 4 distribution amplitudes for
  which we do not include SU(3) breaking. Renormalization scale as in
  previous table.}\label{tab:para2}
\renewcommand{\arraystretch}{1}
\addtolength{\arraycolsep}{-3pt}
\end{table}
The model for the leading twist~2 distribution amplitude $\phi_\perp$ is
\begin{equation}\label{eq:phiperp}
\phi_\perp(u) =  6 u\bar u \left[ 1 + 3 a_1^\perp\, \xi +
a_2^\perp\, \frac{3}{2} ( 5\xi^2  - 1 ) \right]
\end{equation}
with parameter values as specified in Tab.~\ref{tab:para}.
The expressions for higher-twist distributions given below correspond
to the simplest self-consistent approximation which satisfies all 
QCD equations of motion \cite{BBKT,BBS} :
\begin{itemize}
\item{} Three-particle distribution of twist~3 :
\begin{equation}
{\cal T} (\underline{\alpha}) = 
540\, \zeta_3 \omega^T_3 (\alpha_d-\alpha_u)\alpha_d \alpha_u\alpha_g^2.
\end{equation}
\item{} Two-particle distributions of twist~3 :
\begin{eqnarray}
h_\parallel^{(s)}(u) & = & 6u\bar u \left[ 1 + a_1^\perp \xi + \left( \frac{1}{4}a_2^\perp +
\frac{5}{8}\,\zeta_{3}\omega_3^T \right) (5\xi^2-1)\right]\nonumber\\
& & {}+ 3\, \delta_+\, (3 u \bar u + \bar u \ln \bar u + u \ln u) + 
3\,\delta_-\,  (\bar u
\ln \bar u - u \ln u),\label{eq:e}\\
h_\parallel^{(t)}(u) &= & 3\xi^2+ \frac{3}{2}\,a_1^\perp \,\xi (3 \xi^2-1)
+ \frac{3}{2} a_2^\perp\, \xi^2 \,(5\xi^2-3) 
+\frac{15}{16}\zeta_{3}\omega_3^T(3-30\xi^2+35\xi^4)\nonumber\\
& & {} + \frac{3}{2}\,\delta_+
\, (1 + \xi \, \ln \bar u/u) + \frac{3}{2}\,\delta_- \, \xi\, ( 2
+ \ln u + \ln\bar u )
\label{eq:hL}
\end{eqnarray}
\item{} Three-particle distributions of twist~4 :
\begin{eqnarray}
 T^{(4)}_1(\underline{\alpha}) &=& T^{(4)}_3(\underline{\alpha}) ~~=~~0, 
 \nonumber\\
 T^{(4)}_2(\underline{\alpha}) &=& 
    30 \widetilde \zeta^T_{4}(\alpha_d-\alpha_u)\alpha_g^2,
 \nonumber\\
 T^{(4)}_4(\underline{\alpha}) &=& 
    - 30 \zeta^T_{4}(\alpha_d-\alpha_u)\alpha_g^2,
 \nonumber\\
 S(\underline{\alpha}) &=& 30 \zeta^T_{4}(1-\alpha_g)\alpha_g^2,
 \nonumber\\
 \widetilde S(\underline{\alpha}) &=& 
       30 \widetilde \zeta^T_{4}(1-\alpha_g)\alpha_g^2.
\end{eqnarray}
\item{} Two-particle distributions of twist~4 :
\begin{eqnarray}
   h_3(u) &=& 6u(1-u)+5[\zeta^T_4+\widetilde \zeta^T_4](1-3\xi^2),
\nonumber\\
   A^{(o)}(u) &=& 30 u^2(1-u)^2
   \Bigg[\frac{2}{5}+\frac{4}{3}\zeta^T_4-\frac{8}{3}
   \widetilde \zeta^T_4\Bigg].
\end{eqnarray}
\end{itemize}
The constants $\zeta^T_4$ and $\widetilde \zeta^T_4$
are defined as
\begin{eqnarray}
\langle 0|\bar u gG_{\mu \nu}d|\rho^-(P,\lambda)\rangle &=&
  if_\rho^T m_\rho^3 \zeta^T_4(
e^{(\lambda)}_{\mu}P_\nu - e^{(\lambda)}_{\nu}P_\mu),
\nonumber\\
\langle 0|\bar u g\widetilde G_{\mu \nu}i\gamma_5 
   d|\rho^-(P,\lambda)\rangle &=&
  if_\rho^T m_\rho^3 \widetilde \zeta^T_4(
e^{(\lambda)}_{\mu}P_\nu - e^{(\lambda)}_{\nu}P_\mu)
\end{eqnarray}
and have been estimated in \cite{BBK} from QCD sum rules:
\begin{equation}
    \zeta^T_4 \simeq - \widetilde \zeta^T_4 \simeq 0.10.
\end{equation}
Other parameters are given in 
Tab.~\ref{tab:para}\footnote{In notations of Ref.~\cite{BBKT}
$\zeta_3^T \equiv (3/28)\zeta_3\omega_3^T$.}.
Like in  the chiral-even case, SU(3) breaking corrections are neglected
in twist~4 distributions.

The number $\omega_3^T$, $a_1^\perp$, $a_2^\perp$ are computed in the appendix C of \cite{BBKT}, using usual two-points QCD sum rule. $\delta_+$ and $\delta_-$ have the following definition :
\be
\delta_\pm =\frac{f_\rho}{f^T_\rho}\frac{m_u\pm m_d}{m_\rho}
\ee

%
%

\sect{Formulas for Orthogonal Polynomials\label{app:a}}
This appendix reproduces useful formulas for the Appell, Jacobi and Gegenbauer polynomials.

\noindent Differentiation formula for Gegenbauer polynomials :
\begin{equation}
\frac{d}{d\xi} (1-\xi^{2})C_{n}^{3/2}(\xi) = 
-(n+1)(n+2) C_{n}^{1/2}(\xi).
\label{eq:jg1}
\end{equation}
Differentiation formulae for Jacobi polynomials :
\begin{eqnarray}
\frac{d}{d\xi} P_{n}^{(\nu_{1}, \nu_{2})}(\xi)
&=& \frac{n + \nu_{1} + \nu_{2} + 1}{2} 
P_{n-1}^{(\nu_{1}+1,\nu_{2}+1)}(\xi),
\label{eq:jd2}\\
(1 + \xi) P_{n}^{(1,1)}(\xi)
&=&  2\frac{d}{d\xi}\left[
\frac{(n+1)}{(n+2)(2n+3)} P_{n+2}^{(0,0)}(\xi)\right. 
\nonumber \\
& &\left. {}+\frac{1}{n+2} P_{n+1}^{(0,0)}(\xi)
+ \frac{1}{2n+3}P_{n}^{(0,0)}(\xi)\right].
\label{eq:jd1}
\end{eqnarray}
The equation (\ref{eq:jd1}) is obtained from (\ref{eq:jd2})
combined with (\ref{eq:jrec1}) below.

\noindent
Recurrence formulae for Jacobi polynomials :
\begin{eqnarray}
(1+\xi) P_{n}^{(1,1)}(\xi) &=& \frac{(n+1)(n+3)}{(n+2)(2n+3)}
P_{n+1}^{(1,1)}(\xi) + P_{n}^{(1,1)}(\xi) + \frac{n+1}{2n+3}
P_{n-1}^{(1,1)}(\xi)
\label{eq:jrec1}\nonumber \\ \\
&=&\frac{2(n+1)}{2n+3} \left(P_{n}^{(1,0)}(\xi) + P_{n+1}^{(1,0)}(\xi)
\right),
\label{eq:jrec12}\\
(1-\xi) P_{n}^{(1,1)}(\xi) &=&
\frac{2(n+1)}{2n+3} \left(P_{n}^{(0,1)}(\xi) - P_{n+1}^{(0,1)}(\xi)
\right),
\label{eq:jrec13}\\
P_{n}^{(0,0)}(\xi) &=& \frac{n+1}{2n+1}P_{n}^{(1,0)}(\xi)
- \frac{n}{2n+1}P_{n-1}^{(1,0)}(\xi)
\nonumber \\
&=&
\frac{n+1}{2n+1}P_{n}^{(0,1)}(\xi)
+ \frac{n}{2n+1}P_{n-1}^{(0,1)}(\xi),
\label{eq:jrec2}
\end{eqnarray}
\beq
P_n^{(0,0)}(\xi) + P_{n+1}^{(0,0)}(\xi) &=&
(1+\xi)P^{(0,1)}_n(\xi),
\label{A1}\\
P_n^{(0,0)}(\xi) - P_{n+1}^{(0,0)}(\xi) &=& 
(1-\xi)P^{(1,0)}_n(\xi). 
\label{A2}
\eeq
Relations between Jacobi and Gegenbauer polynomials :
\begin{eqnarray}
(1+\xi) P_{n}^{(0,1)}(\xi) + (1-\xi)P_{n}^{(1,0)}(\xi) &=&
2 C_{n}^{1/2}(\xi),
\label{eq:jg2}\\
(1+\xi) P_{n}^{(0,1)}(\xi) - (1-\xi)P_{n}^{(1,0)}(\xi) &=&
2 C_{n+1}^{1/2}(\xi),
\label{eq:jg3}
\end{eqnarray}
\begin{equation}
(n+2)P_{n}^{(1,1)}(\xi) = 2 C_{n}^{3/2}(\xi).
\label{eq:jg4}
\end{equation}
Orthogonality relations for Appell polynomials~\cite{Er} :
\begin{equation}
\int{\cal D}\underline{\alpha} \: \alpha_{d}\alpha_{u}\alpha_{g}^{2}
J_{k,l}(\alpha_{d}, \alpha_{u})
J_{m,n}(\alpha_{d}, \alpha_{u})
= \delta_{k+l, m+n}
\frac{(-1)^{k+l}}{2^{k+l+3}(k+l+3)(2k+2l+5)!!}
W_{k,m}^{(k+l+1)},
\label{eq:appello0}
\end{equation}
where $W_{k,m}^{(k+l+1)} \equiv 
\partial^{m+n}J_{k,l}(\alpha_{d}, \alpha_{u})/
\partial \alpha_{d}^{m} \partial \alpha_{u}^{n}$
is a $(k+l+1)\times (k+l+1)$ symmetric matrix. This result can be
obtained from the following relations:
\begin{eqnarray}
\lefteqn{\int {\cal D}
\underline{\alpha}\: \alpha_{d}^{m+1}\alpha_{u}^{n+1}
\alpha_{g}^{2} J_{k,l}(\alpha_{d}, \alpha_{u}) =}\makebox[2cm]{\ }
\nonumber \\
&=&\left\{
\begin{array}{@{\,}ll}
0& (m+n < k+l)\\
\delta_{m,k} 
\displaystyle{\frac{(-1)^{k+l}k!l!}{2^{k+l+3}(k+l+3)(2k+2l+5)!!}}&
(m+n=k+l),
\end{array}
\right.
\label{eq:appello}
\end{eqnarray}
while the integral is in general nonzero for $m+n>k+l$.

\noindent
Integral formulae for Appell polynomials :
\begin{eqnarray}
&\mbox{$ $}&\frac{d}{du}\int_{0}^{u}\!d\alpha_{d} \int_{0}^{\overline{u}}
\!d\alpha_{u}\, 
\frac{1}{1 -\alpha_{d}-\alpha_{u}}\left(
\alpha_{d}\frac{\partial}{\partial \alpha_{d}}
+\alpha_{u}\frac{\partial}{\partial \alpha_{u}}
- 1\right) \alpha_{d} \alpha_{u} (1 - \alpha_{d}-\alpha_{u})^{2}
J_{k,l}(\alpha_{d}, \alpha_{u}) =
\nonumber\\
&\mbox{$ $}&\;\;\;\;\;\;\;\;\;\;\;\;
= \frac{u\overline{u}}{2}\frac{k!l!(-1)^{k}}{(k+l+2)!}
\left(\frac{k-l}{k+l+3}P_{k+l+2}^{(1,1)}(\xi)
+ P_{k+l+1}^{(1,1)}(\xi)\right),
\label{eq:apint}
\end{eqnarray}
\begin{eqnarray}
\lefteqn{\frac{d}{du}\int_{0}^{u}d\alpha_{d} \int_{0}^{\overline{u}}
d\alpha_{u} 
\alpha_{d} \alpha_{u} (1 - \alpha_{d}-\alpha_{u})
J_{k,l}(\alpha_{d}, \alpha_{u}) =}\makebox[1cm]{\ }
\nonumber\\
&=& \frac{u\overline{u}}{2}\frac{k!l!(-1)^{k}}{(k+l+3)!}
\left(\frac{k-l}{k+l+3}P_{k+l+2}^{(1,1)}(\xi)
- P_{k+l+1}^{(1,1)}(\xi)\right),
\label{eq:apint2}
\end{eqnarray}
\beq
& &{d\over du}
\int_0^ud\alpha_d \int_0^{\bar{u}}d\alpha_u {1\over 1- \alpha_d -\alpha_u}
\left( \alpha_d {\partial \over \partial\alpha_d } 
- \alpha_u{\partial\over \partial\alpha_u}\right)
\alpha_d\alpha_u (1-\alpha_d-\alpha_u)^2 J_{k,l}(\alpha_d,\alpha_u)\ =
\nonumber\\
& &\qquad ={u\bar{u}\over 2}{k!l!(-1)^k\over (k+l+3)!} 
\left[{(k+l+2)(k+l+4)\over k+l+3}
P^{(1,1)}_{k+l+2}(\xi)
+(k-l)P^{(1,1)}_{k+l+1}(\xi) \right].
\label{A3}
\eeq
\beq
& &\int_0^ud\alpha_d \int_0^{\bar{u}}d\alpha_u {1\over 1- \alpha_d -\alpha_u}
\left( {\partial\over \partial\alpha_d } 
+ {\partial\over \partial\alpha_u}\right)
\alpha_d\alpha_u (1-\alpha_d-\alpha_u)^2 J_{k,l}(\alpha_d,\alpha_u)\ =
\nonumber\\
& &\qquad ={k!l!(-1)^k\over (k+l+3)!4} 
\left[{-k+l\over 2k+2l+5}
\left( 
P^{(0,0)}_{k+l+1}(\xi)-
P^{(0,0)}_{k+l+3}(\xi) \right)\right. \nonumber\\
& &\left.\qquad\qquad\qquad 
+{k+l+2\over 2k+2l+7}
\left( 
P^{(0,0)}_{k+l+2}(\xi)-
P^{(0,0)}_{k+l+4}(\xi) \right)\right].
\label{A4}
\eeq
The results (\ref{eq:apint})$-$(\ref{A3}) 
can be obtained by differentiating and/or
integrating 
the Appell polynomials $J_{k,l}(\alpha_{d}, \alpha_{u})$ term by term.
To obtain (\ref{A4}), it is convenient to calculate its derivative
first, which can be done similarly to (\ref{eq:apint}), (\ref{eq:apint2})
and (\ref{A3}),   
and then integrate the result with the condition that it vanishes
 at $u=0$.


\sect{Operators relations for different theoretical twist \label{app:oprel}}

This appendix contains some relations taken from \cite{BaBr}, \cite{BB96} and \cite{BBK}, which help to isolate the different twist parts of two-points non-local operators.

\subsection{Chiral-even operator}

Consider the non-local operators $\ov{\psi}(x)\gamma_\alpha\psi(-x)$. The local expansion around $x=0$ can be written in the following way :
\be
\ov{\psi}(x)\gamma_\alpha\psi(-x)=\sum_{n=0}^\infty x_{\mu_1}\ldots x_{\mu_n}
\frac{1}{n!}\ov{\psi}\Drl_{\mu_1}\ldots\Drl_{\mu_n}\gamma_\alpha\psi
\ee

In order to get the leading twist contribution, the symmetrization has to be done over all indices and subtract the traces :
\be
\left[\ov{\psi}(x)\gamma_\alpha\psi(-x)\right]_{\text{l. t.}}\equiv\sum_{n=0}^\infty x_{\mu_1}\ldots x_{\mu_n}
\frac{1}{n!}\ov{\psi}\left(\Drl_{\mu_1}\ldots\Drl_{\mu_n}\gamma_\alpha\right)
_{\text{sym}}\psi-\text{traces}
\ee

The symmetrization has an integrated solution :
\begin{eqnarray}
\lefteqn{\Big[\bar \psi(x)\gamma_\alpha \psi(-x)\Big]_{\text{sym}} \equiv}
\nonumber\\
&\equiv &
\sum_{n=0}^\infty \frac{x^{\mu_1}\ldots x^{\mu_n}}{n!}
\bar \psi(0)\Bigg\{\frac{1}{n+1}\Drl_{\mu_1}\ldots \Drl_{\mu_n}\gamma_\alpha 
+\frac{n}{n+1}\Drl_{\alpha}\Drl_{\mu_1}\ldots \Drl_{\mu_{n-1}}\gamma_{\mu_n}
\Bigg\} \psi(0)\,.\makebox[1cm]{\ } \nonumber \\
\label{defsym} \\
 & =&
\int_0^1 dv\,\frac{\partial}{\partial x_\alpha}\bar \psi(vx)\hat{x} 
\psi(-v x)
\label{BBformula}
\eq
with $\hat{x}\equiv\gamma_\mu x^\mu$.

In order to have really the leading twist part of  $\ov{\psi}(x)\gamma_\alpha\psi(-x)$, the leading twist part of $\bar \psi(x)\hat{x} \psi(-x)$ is needed. If one does a local expansion of this latter non-local operator, one will already have symetrized local operators. But traces has alos to be removed; this can be expressed as a differential equation :
\be
\frac{\partial}{\partial x_\alpha\partial x^\alpha}\left[\ov{\psi}(x)\hat{x}
\psi(-x)\right]_{\text{l. t.}}=0 \label{ltxhatapp}
\ee
which has the formal solution :
\bq
\left[\ov{\psi}(x)\hat{x}\psi(-x)\right]_{\text{l. t.}}&=&\ov{\psi}(x)\hat{x}\psi(-x)
\nonumber \\
& &{}+\sum_{n=1}^\infty\int_0^1\left(-\frac{1}{4}x^2\right)^n
\frac{\left(\frac{1-t}{t}\right)^{n-1}}{(n-1)!n!}
\left(\frac{\partial^2}{\partial x^\alpha\partial x_\alpha}\right)^n
\ov{\psi}(tx)\hat{x}\psi(-tx) \nonumber \\ \label{solltxhat}
\eq

Equations for the twist 3 part of $\ov \psi(x)\gamma_\alpha\psi(-x)$  are the following :

\begin{eqnarray}
 \Big[\bar \psi(x)\gamma_\mu \psi(-x)\Big]_{\text{twist 3}} &=&
{}-g_s\int_0^1 \!du\int_{-u}^u \!dv\,\bar \psi(ux)
\left[
u\tilde G_{\mu\nu}(vx)x^\nu \hat{x}\gamma_5\right. \nonumber \\ 
& &{} -ivG_{\mu\nu}(vx)x^\nu \hat{x}
\Big]\psi(-ux)
\nonumber\\
&&{}+i\epsilon_{\mu}^{\phantom{\mu}\nu\alpha\beta}\int_0^1 udu\, 
x_\nu\partial_\alpha
\Big[\bar \psi(ux)\gamma_\beta\gamma_5 \psi(-ux)\Big]\,,
\nonumber\\
\Big[\bar \psi(x)\gamma_\mu\gamma_5 \psi(-x)\Big]_{\text{twist 3}} &=&
-g_s\int_0^1 \!du\int_{-u}^u \!dv\,\bar \psi(ux)
\left[
u\tilde G_{\mu\nu}(vx)x^\nu\hat{x}\right. \nonumber \\ & &{}-ivG_{\mu\nu}(vx)x^\nu\hat{x}\gamma_5
\Big]\psi(-ux)
\nonumber\\
&&{}+i\epsilon_{\mu}^{\phantom{\mu}\nu\alpha\beta}\int_0^1 udu\, 
x_\nu\partial_\alpha
\Big[\bar \psi(ux)\gamma_\beta \psi(-ux)\Big]\,,
\label{twist3}
\end{eqnarray}
where $G_{\mu\nu}$ is the gluon field strength, $\tilde G_{\mu\nu}
=(1/2)\epsilon_{\mu\nu\alpha\beta}G^{\alpha\beta}$, and
$\partial_\alpha$ is the derivative over the total translation :
\begin{equation}
\partial_\alpha\Big[\bar \psi(ux)\gamma_\beta \psi(-ux)\Big] \equiv 
\left.\frac{\partial}{\partial y_\alpha}
\Big[\bar \psi(ux+y)\gamma_\beta \psi(-ux+y)\Big]\right|_{y\to 0}. \label{totder} 
\end{equation}

These equations come from \cite{BB96} and can be obtained with the equation of motion, neglecting quark masses.

Another useful formula is the equation (5.13) in \cite{BaBr} which can be written like this :
\be
\frac{\partial}{\partial x_\alpha\partial x^\alpha}\ov \psi(x)\hat{x}\psi(-x)=
-\partial^2\ov \psi(x)\hat{x}\psi(-x) + \text{operators with gluons}
\ee

The expansion of \ref{solltxhat} at the order $x^2$, gives
\bq
\ov \psi(x)\hat{x}\psi(-x)&=&\left[\ov \psi(x)\hat{x}\psi(-x)\right]_
{\text{l. t.}}+\frac{1}{4}x^2\int_0^1dv\frac{\partial^2}{\partial x_\alpha
\partial x^\alpha}\ov \psi(vx)\hat{x}\psi(-vx) + \ldots \nonumber \\ 
&=&\left[\ov \psi(x)\hat{x}\psi(-x)\right]_
{\text{l. t.}}-\frac{1}{4}x^2\int_0^1dv v^2\partial^2
\ov \psi(vx)\hat{x}\psi(-vx) \nonumber \\
& &{}+ \text{operators with gluons}+\text{twist 3}+\ldots \label{dertot}
\eq

This last equation gives a twist 4 contribution with a total derivative (defined in \ref{totder}).

\subsection{Chiral-odd operator} 

Consider the non-local operators $\ov{\psi}(x)\sigma_{\alpha\beta}\psi(-x)$. To get the most symmetrical part in the local expansion, one can compare it with the solution in \ref{BBformula} and get directly :
\begin{eqnarray}
\lefteqn{\Big[\bar \psi(x)\sigma_{\alpha\beta} \psi(-x)\Big]_{\text{sym}} =} \nonumber
\\
 & =&
\int_0^1 dv\,\frac{\partial}{\partial x_\beta}\bar \psi(vx)
\sigma_{\alpha\nu}x^\nu 
\psi(-v x)-(\alpha \leftrightarrow \beta)
\label{BBformulasig}
\eq

In \cite{BBK}, the equations for the leading twist part of $\bar \psi(x)\sigma_{\alpha\nu}x^\nu \psi(-x)$ can be founds~:
\bq
\frac{\partial}{\partial x_\alpha}\left[\bar \psi(x)\sigma_{\alpha\nu}x^\nu \psi(-x)\right]_{\text{l. t.}}&=&0 \\
\frac{\partial^2}{\partial x_\rho \partial x^\rho}\left[\bar \psi(x)\sigma_{\alpha\nu}x^\nu \psi(-x)\right]_{\text{l. t.}}&=&0
\eq

Like in equation \ref{dertot}, the contribution of total derivative can be added :
\bq
\bar \psi(x)\sigma_{\alpha\nu}x^\nu \psi(-x)&=&
\left[\bar \psi(x)\sigma_{\alpha\nu}x^\nu \psi(-x)\right]_{\text{l. t.}} -\frac{1}{4} x^2x^\nu\int_0^1 dt\,t^2 \partial^2\left(\bar \psi(tx)\sigma_{\alpha\nu}\psi(-tx)\right) \nonumber \\
& &{}+\text{operators with gluons}+\ldots 
\eq


\sect{Mass corrections to two-points distributions \label{app:masscorr}}

In this appendix, the mass corrections for the distributions are given, using the method described in section \ref{thtmasscorr}.

\subsection{Chiral even distributions}

For the different wave functions
\begin{eqnarray}
\lefteqn{\langle 0|\bar u(x) \gamma_{\mu} [x,-x] 
d(-x)|\rho^-(P,\lambda)\rangle = }\makebox[2cm]{\ } \nonumber \\
&=& f_{\rho} m_{\rho} \left[ P_{\mu}
\frac{e^{(\lambda)}\cdot x}{P \cdot x}
\int_{0}^{1} \!du\, e^{i \xi P \cdot x} \left(A^{(e)}(u)+
\frac{x^2 m_\rho^2}{4}\hat{A}^{(e)}(u)\right) \right. 
\nonumber \\
& &{}+ \left(e^{(\lambda)}_{\mu}-P_\mu\frac{\epl\cdot x}{P\cdot x}\right)
\int_{0}^{1} \!du\, e^{i \xi P \cdot x} \left(B^{(e)}(u) 
+\frac{x^2 m_\rho^2}{4}\hat{B}^{(e)}(u)\right)\nonumber \\
& & \left.{}- \frac{1}{2}x_{\mu}
\frac{e^{(\lambda)}\cdot x }{(P \cdot x)^{2}} m_{\rho}^{2}
\int_{0}^{1} \!du\, e^{i \xi P \cdot x} C^{(e)}(u) \right]
\end{eqnarray}
one has
\bq
\int_0^1duB^{(e)}(u)e^{i\xi Px}&=&\int_0^1du A{(e)}(u)\int_0^1dte^{it\xi Px}\nonumber
\\
& &{}-(P\cdot x)^2\int_0^1du\int_0^1dt\,t\int_0^1dv\,v^2e^{ivt\xi Px}B^{(e)}(u)
\nonumber \\ \\
-\frac{1}{2}\frac{1}{(P\cdot x)^2}\int_0^1duC^{(e)}(u)e^{i\xi Px}&=&\int_0^1du\frac{\xi^2+1}{\xi}A^{(e)}(u)\int_0^1dt\,t^2\int_0^1dvv^2e^{ivt\xi Px} \nonumber \\ 
& &{}-\int_0^1du B^{(e)}(u)\int_0^1dt\,t\int_0^1dv\,v^2e^{ivt\xi Px} \nonumber \\
\\
\int_0^1du\hat{B}^{(e)}(u)e^{i\xi Px}&=&
\int_0^1duA^{(e)}(u)(\xi^2+1)\int_0^1dt\,t^2\int_0^1v^2dve^{ivt\xi Px} \nonumber \\
& &{}+4\int_0^1duB^{(e)}(u)\int_0^1dt\,t\int_0^1dv\,v^2e^{ivt\xi Px} \nonumber
\\
& &{}-(P\cdot x)^2\int_0^1du\hat{B}^{(e)}(u)\int_0^1dt\,t^3
\int_0^1dv\,v^3e^{itv\xi Px} \nonumber \\ 
\eq
\bq\lefteqn{\int_0^1du\hat{A}^{(e)}(u)e^{i\xi Px}=} \nonumber \\
& &{}
\int_0^1duA^{(e)}(u)(\xi^2+1)\int_0^1dt\,t^2\int_0^1dv\,v^2\left(i\xi(P\cdot x)+1\right)
t^2e^{ivt\xi Px} \nonumber \\
& &{}+4\int_0^1duB^{(e)}(u)\int_0^1dt\,t\int_0^1dv\,v^2e^{ivt\xi Px} 
\nonumber \\
\eq

\subsection{Chiral odd distributions}
For the different wave functions
\begin{eqnarray}
\lefteqn{\hspace*{-1.5cm}\langle 0|\bar u(x) \sigma_{\mu \nu} [x,-x] 
d(-x)|\rho^-(P,\lambda)\rangle =} \nonumber \\
&=& i f_{\rho}^{T} \left[ ( e^{(\lambda)}_{\mu}P_\nu -
e^{(\lambda)}_{\nu}P_\mu )
\int_{0}^{1} \!du\, e^{i \xi P \cdot x} \left(A^{(o)}(u) 
+\frac{m^2_\rho x^2}{4}\hat{A}^{(o)}(u)\right)\right. 
\nonumber \\
& &{}+ (P_\mu x_\nu - P_\nu x_\mu )
\frac{e^{(\lambda)} \cdot x}{(P \cdot x)^{2}}m_{\rho}^{2} 
\int_{0}^{1} \!du\, e^{i \xi P \cdot x}\left(B^{(o)}(u) 
+\frac{m^2_\rho x^2}{4}\hat{B}^{(o)}(u)\right)  
\nonumber \\
& & \left.{}+ \frac{1}{2}
(e^{(\lambda)}_{\mu} x_\nu -e^{(\lambda)}_{\nu} x_\mu) 
\frac{m_{\rho}^{2}}{P \cdot x} 
\int_{0}^{1} \!du\, e^{i \xi P \cdot x} C^{(o)}(u) \right],
\end{eqnarray}
one has
\bq
\lefteqn{\int_0^1due^{i\xi Px}\hat{A}^{(o)}(u)=} \nonumber \\
&=&\int_u^1duE(u)\int_0^1dt\int_0^1dv\,v^3e^{ivt\xi Px}
\left[-\frac{2}{tv^3(Px)^3}
-\frac{2}{v^3(Px)^3}+2\frac{i\xi}{v^2(Px)^2}\right. \nonumber \\
& &{}\left.+2\frac{i\xi t}{v^2(Px)^2}
-3\frac{(i\xi)^2t}{v(Px)}+2\frac{i\xi)^2t^2}{v(Px)}-(i\xi)^3t^2+\frac{t}{v(Px)}
+t^2(i\xi)\right] \nonumber \\
\eq
\bq
\lefteqn{\int_0^1due^{i\xi Px}C^{(o)}(u)=} \nonumber \\
&=&\frac{1}{2}(P\cdot x)\int_0^1duE(u)\int_0^1dt\int_0^1dv\,v^2e^{itv\xi Px}
\left[\frac{4}{tv^2(Px)}-4\frac{i\xi}{v(Px)}-2(i\xi)^2t+2t\right]\nonumber \\
\eq
\bq
\lefteqn{\int_0^1due^{i\xi Px}B^{(o)}(u)=} \nonumber \\
&=&(P\cdot x )\int_0^1duE(u)\int_0^1dt\int_0^1dv\,v^2e^{itv\xi Px}
\left[-\frac{1}{tv^2(Px)}+\frac{1}{v^2(Px)}+\frac{i\xi}{v(Px)}\right.\nonumber \\
& &{}-\left.
\frac{i\xi t}{v(Px)}+\frac{(i\xi)^2t}{2}-(i\xi)^2t^2-\frac{t}{2}\right]
\eq
where
\be
E(u)\equiv\frac{1}{2}i\frac{d}{du}A^{(o)}(u)
\ee

There is no equation for $\hat{B}^{(e)}(u)$ because it is already a $O(m^4_\rho)$ contribution.

\end{appendix}

\end{document}